\documentclass[journal,twocolumn,twoside]{IEEEtran}

\UseRawInputEncoding

\usepackage{graphics} 
\usepackage{amssymb}  
\usepackage{cite}
\usepackage{amsmath}
\usepackage{algorithm}
\usepackage{algorithmic}
\usepackage{graphicx}
\usepackage{epstopdf}
\usepackage{booktabs}
\usepackage{multirow}
\usepackage{etoolbox}
\usepackage{mathdots}
\usepackage{xcolor}

\makeatletter
\patchcmd{\@makecaption}
  {\scshape}
  {}
  {}
  {}
\makeatother
\usepackage{diagbox}
\usepackage{bm}
\usepackage{subfigure}

\def\norm #1{\left\|#1\right\|}

\def\twon #1{\left\|#1\right\|_2}
\def\onen #1{\left\|#1\right\|_1}

\def\frobn #1{\left\|#1\right\|_{\text{F}}}

\def\abs #1{\left|#1\right|}
\def\inp #1{\left\langle#1\right\rangle}

\def\st{\text{subject to }}

\def\bC{\mathbb{C}}

\def\bR{\mathbb{R}}
\def\bE{\mathbb{E}}

\def\m #1{\boldsymbol{#1}}

\def\cC{\mathcal{C}}

\def\cE{\mathcal{E}}

\def\cH{\mathcal{H}}
\def\cI{\mathcal{I}}

\def\cP{\mathcal{P}}

\def\cT{\mathcal{T}}

\def\cZ{\mathcal{Z}}

\def\bee{\begin{equation}}
\def\ene{\end{equation}}

\def\beq{\begin{eqnarray}}
\def\enq{\end{eqnarray}}
\def\lentwo{\setlength\arraycolsep{2pt}}

\newtheorem{lem}{Lemma}

\newtheorem{thm}{Theorem}
\newtheorem{prop}{Proposition}

\newenvironment{proof}{{\noindent\it \quad Proof:}\;}{\hfill $\square$\par}

\def\equ #1{\begin{equation}#1\end{equation}}
\def\equa #1{\begin{eqnarray}#1\end{eqnarray}}
\def\sbra #1{\left(#1\right)}
\def\mbra #1{\left[#1\right]}
\def\lbra #1{\left\{#1\right\}}
\def\diag #1{\text{diag}#1}
\def\tr #1{\text{tr}#1}

\def\rank #1{\text{rank}#1}
\def\st {\text{ subject to }}

\title{New Low Rank Optimization Model and Convex Approach for Robust Spectral Compressed Sensing}
\author{Zai Yang and Xunmeng Wu
\thanks{Part of this paper will be presented at the 2020 European Signal Processing Conference (EUSIPCO) \cite{yang2020forward}.

The authors are with the School of Mathematics and Statistics, Xi'an Jiaotong University, Xi'an 710049, China (e-mail: yangzai@xjtu.edu.cn).}}

\begin{document}
\maketitle

\begin{abstract}
This paper investigates recovery of an undamped spectrally sparse signal and its spectral components from a set of regularly spaced samples within the framework of spectral compressed sensing and super-resolution. We show that the existing Hankel-based optimization methods suffer from the fundamental limitation that the prior of undampedness cannot be exploited. We propose a new low rank optimization model partially inspired by forward-backward processing for line spectral estimation and show its capability in restricting the spectral poles on the unit circle. We present convex relaxation approaches with the model and show their provable accuracy and robustness to bounded and sparse noise. All our results are generalized from the 1-D to arbitrary-dimensional spectral compressed sensing. Numerical simulations are provided that corroborate our analysis and show efficiency of our model and advantageous performance of our approach in improved accuracy and resolution as compared to the state-of-the-art Hankel and atomic norm methods.

\end{abstract}



\section{Introduction}
Spectral compressed sensing \cite{candes2006robust,duarte2013spectral} refers to the recovery or estimation of a spectrally sparse signal and its spectral components from a subset of regularly spaced samples. In particular, the regularly spaced samples $\lbra{\tilde y_j}$ are given by
\equ{\tilde y_j=y_j^o+e_j, \quad y_j^o = \sum_{k=1}^K s_k z_k^{j-1},\quad \abs{z_k}=1, \label{eq:model}}
where $y_j^o$ denotes the ground truth signal, $e_j$ is noise, and $z_k,s_k$ denote the pole and complex amplitude of the $k$th spectral component. The signal is called spectrally sparse since the number of spectral components $K$ is small. In this paper, we consider undamped spectral signals, which are commonly encountered in array signal processing, radar, wireless communications, and many other areas \cite{stoica2005spectral}, with $z_k = e^{i2\pi f_k}$, $k=1,\dots,K$ lying on the unit circle that have a one-to-one connection to the frequencies $f_k\in[0,1)$, $k=1,\dots,K$ where $i=\sqrt{-1}$. Our objective is to estimate $\lbra{y_j^o}$, $\lbra{z_k}$ and $\lbra{s_k}$ given partial entries in $\lbra{\tilde y_j}$.

Spectral compressed sensing is known also as off-the-grid compressed sensing \cite{tang2013compressed} in the sense that the frequencies in the model are not assumed on a fixed grid as opposed to conventional compressed sensing \cite{candes2006robust}. When all entries in $\lbra{\tilde y_j}$ are available, it is the well-known line spectral estimation problem \cite{stoica2005spectral} that is fundamental in digital signal processing (where the unobserved samples are called missing data). The latter case is known also as spectral super-resolution for emphasizing the extent to which nearly located frequencies can be resolved \cite{candes2013towards,candes2013super}.

Line spectral estimation (in the case when all regularly spaced samples are available) has a long history of research, and a great number of approaches have been developed, see \cite{stoica2005spectral}. Among them, the most well-known approach is the (nonparametric) fast Fourier transform (FFT), one of the top ten algorithms in the 20th century \cite{dongarra2000guest}. With the rapid increase of computing power and the need of higher accuracy and resolution, parametric approaches like maximum likelihood estimation (MLE) were developed in which the main difficulty comes from the nonlinearity and nonconvexity with respect to the frequency parameters. To overcome this difficulty, subspace-based methods, e.g.~MUSIC and ESPRIT \cite{schmidt1981signal,roy1986esprit}, were proposed that turn to estimate the frequencies from the second-order statistics estimated from the samples. The subspace methods have good statistical properties \cite{fannjiang2016compressive,liao2016music}; however, they need the model order $K$ that is usually unknown and have poor performance in the presence of missing data and outliers.

Sparse optimization and compressed sensing approaches have dominated the research of this century on this topic, see \cite{yang2018sparse}. By exploiting the spectral sparsity and proposing an optimization framework for signal recovery, the aforementioned drawbacks of subspace methods can be tackled. The idea of sparse methods for spectral compressed sensing dates back to the last century \cite{gorodnitsky1997sparse}. But rigourous theory and algorithms have not been established until the seminal papers \cite{candes2013towards,candes2013super} in which a convex optimization approach was proposed to deal with the continuous-valued frequencies. To date, such convex, {\em gridless} sparse methods include atomic norm (or total variation norm) methods \cite{chandrasekaran2012convex, candes2013towards,bhaskar2013atomic,tang2013compressed,azais2015spike, yang2016vandermonde}, enhanced matrix completion (EMaC) \cite{chen2014robust} and covariance fitting approaches \cite{yang2015gridless,wu2017toeplitz,zhou2018direction}. Unlike the conventional MLE that solves directly for the frequencies in a nonconvex way, interestingly, all these methods turn to optimize explicitly or implicitly a structured low rank matrix where the low-rankness comes from the spectral sparsity. In particular, atomic norm and covariance fitting are related to a low rank positive semidefinite (PSD) Toeplitz matrix, and EMaC is rooted on a low rank Hankel matrix. Consequently, spectral compressed sensing is connected to the celebrated low rank matrix recovery problem \cite{davenport2016overview}.

\subsection{Main Contributions of This Paper}
We note in this paper that the low rank Hankel model widely employed for spectral compressed sensing has not been well understood. The first contribution of this paper is to show that the low rank Hankel model has a fundamental limitation, to be specific, it cannot use the on-circle prior $\abs{z_k}=1$ in the data model \eqref{eq:model} and its produced poles do not lie on the unit circle in general, resulting in difficulties of physical interpretation and potential performance loss.

The second contribution of this paper is to resolve the limitation above. To do so, we propose a new low rank optimization model for spectral compressed sensing that we call low rank double Hankel model by introducing another Hankel matrix into the model. We show analytically that the new model substantially shrinks the solution space and has great potential to restrict the poles on the unit circle.

By employing the double Hankel model, as the third contribution, we propose new convex optimization approaches to spectral compressed sensing, called doubly enhanced matrix completion (DEMaC) in honor of EMaC, and present theoretical results showing DEMaC's accuracy with full and compressive/partial sampling and its robustness to bounded and sparse noise.

We also extend the double Hankel model and the DEMaC approaches from the 1-D to arbitrary-dimensional spectral compressed sensing. Extensive numerical simulations are provided that confirm our analysis and show that DEMaC performs consistently better than EMaC in various scenarios.

\subsection{Related Work}
It is always an essential and challenging task to exploit the prior $\abs{z_k}=1$ for line spectral estimation. The MLE explicitly uses the prior and suffers from convergence (to global optima) issue. Subspace methods consist of two consecutive estimation steps, first for the (signal or noise) subspace and second for the spectral poles. The forward-backward processing technique \cite{evans1982application,williams1988improved,wang19982} uses the prior in the first step to improve the estimation accuracy of the covariances from which the subspace is computed and eventually improves the accuracy of frequency estimation. MUSIC uses the prior in the second step and performs a line search for frequency estimation \cite{schmidt1981signal}. Besides including the forward-backward technique, unitary ESPRIT also uses the prior in the second step by solving for the spectral poles from a total least squares problem \cite{haardt1995unitary}. Interestingly, root-MUSIC does not use the prior but has higher resolution than MUSIC \cite{barabell1983improving}, which implies that exploiting the prior in the second step might not improve the accuracy because there is no guarantee that the information lost in the first step can be compensated in the second.

The low rank double Hankel optimization model proposed in this paper is partially inspired by the forward-backward processing technique. Unlike subspace methods and as many other optimization approaches, the estimation in DEMaC is accomplished in a single step by solving the resulting optimization problem in which the prior is implicitly included.

Equipped with the low rank Hankel model, nonconvex optimization algorithms have been proposed for spectral compressed sensing \cite{andersson2014new,cho2016fast,cai2019fast}. But these methods need the model order $K$ like the MLE and subspace methods and few theoretical results are known about their accuracy, especially in the practical noisy case. In this paper, we mainly study convex relaxation approaches that do not need the model order and whose accuracy are proved in absence of noise and with bounded and sparse noise. In fact, an algorithm in \cite{cai2019fast} is modified and applied in this paper to the proposed double Hankel model to validate the model efficiency.

\subsection{Notation and Organization}
Notations used in this paper are as follows. The set of real and complex numbers are denoted $\bR$ and $\bC$ respectively. Boldface letters are reserved for vectors and matrices. The amplitude of scalar $a$ is denoted $\abs{a}$. The complex conjugate, transpose and complex transpose of matrix $\m{A}$ are denoted $\overline{\m{A}}$, $\m{A}^T$ and $\m{A}^H$ respectively. The rank of matrix $\m{A}$ is denoted $\rank\sbra{\m{A}}$. We write $\m{A}\geq \m{0}$ if $\m{A}$ is Hermitian and positive semidefinite. The $j$th entry of vector $\m{x}$ is $x_j$. The diagonal matrix with vector $\m{x}$ on the diagonal is denoted $\diag\sbra{\m{x}}$. The Kronecker and Khatri-Rao (a.k.a.~columnwise Kronecker) products of matrices are denoted $\otimes$ and $\star$, respectively.


The rest of this paper is organized as follows. Section \ref{sec:oldmodel} introduces the previous low rank PSD-Toeplitz and Hankel models for spectral compressed sensing. Section \ref{sec:FB-Hankel} presents the new low rank double Hankel model and shows its properties as compared to the Hankel and PSD-Toeplitz models. Section \ref{sec:demac} presents convex relaxation approaches for spectral compressed sensing by applying the double Hankel model and shows their theoretical guarantees with and without noise. Section \ref{sec:simulation} provides numerical results that validate efficiency of the double Hankel model and advantageous performance of DEMaC as compared to EMaC and atomic norm methods. Conclusions are drawn in Section \ref{sec:conclusion} and some technical proofs for DEMaC are given in the Appendix.

\section{Previous Low Rank Optimization Models} \label{sec:oldmodel}
We present previous low rank Hankel and Toeplitz optimization models in this section. Here we assume that the number of spectral poles $K$ is given. Let
\equ{S_0 = \lbra{\m{y}\in\bC^{N}:\; y_j = \sum_{k=1}^K s_k z_k^{j-1},\, s_k\in\bC,\, \abs{z_k}=1} \label{eq:S0}}
that denotes the set of candidate signals that are undamped spectrally sparse with at most $K$ spectral components. What the framework of spectral compressed sensing does is to find some candidate signal $\hat{\m{y}} \in S_0$ that is nearest to $\tilde{\m{y}}=\mbra{\tilde y_1,\dots,\tilde y_N}^T$ and then retrieve the estimated poles from $\hat{\m{y}}$. To be specific, it attempts to solve the following optimization problem:
\equ{\min_{\m{y}} \text{Loss}\sbra{\m{y},\tilde{\m{y}}}, \st \m{y}\in S_0, \label{eq:genprb}}
where $\text{Loss}\sbra{\cdot,\cdot}$ denotes a loss function that is chosen according to the sampling scheme and noise properties. To cast \eqref{eq:genprb} as a tractable optimization problem, the key underlies how to characterize the constraint $\m{y}\in S_0$.

\subsection{Low Rank PSD-Toeplitz Model}
Let
\equ{\cT\m{t} = \begin{bmatrix} t_1 & t_2 & \dots & t_N \\ \overline{t}_2 & t_1 & \cdots & t_{N-1} \\ \vdots & \vdots & \ddots & \vdots \\ \overline{t}_N & \overline{t}_{N-1} & \dots & t_1 \end{bmatrix}}
denote an $N\times N$ Hermitian Toeplitz matrix formed by using $\m{t}\in\bC^N$. By exploiting the Carath\'eodory-Fej\'er theorem \cite[Ch.~4.9.2]{stoica2005spectral}, which states that a PSD Toeplitz matrix admits a Vandermonde decomposition, $S_0$ can be written equivalently as \cite{tang2013compressed,yang2016vandermonde}
\equ{\lbra{\m{y}\in\bC^N:\; \begin{bmatrix} t_1 & \m{y}^H \\ \m{y} & \cT\m{t} \end{bmatrix} \geq \m{0}, \; \rank\sbra{\cT\m{t}}\leq K,\; \m{t}\in\bC^N }. \label{eq:toeplitz_model}}
The PSDness of $\begin{bmatrix} t_1 & \m{y}^H \\ \m{y} & \cT\m{t} \end{bmatrix}$ implies that $\cT\m{t}$ is also PSD. The resulting PSD-Toeplitz model and its variants lead to the atomic norm and covariance fitting methods.

\subsection{Low Rank Hankel Model}
The low rank Hankel model is introduced in \cite{chen2014robust,andersson2014new} inspired by the classical matrix pencil approach \cite{hua1990matrix,hua1992estimating}. For $\m{y}\in \bC^N$, we form the $N_1\times N_2$ Hankel matrix
\equ{\cH\m{y}= \begin{bmatrix} y_{1} & y_{2} & \dots & y_{N_2} \\ y_{2} & y_{3} & \dots & y_{N_2+1} \\ \vdots & \vdots & \ddots & \vdots \\ y_{N_1} & y_{N_1+1} & \dots & y_{N}\end{bmatrix} , \label{eq:Hankel}}
where $N_1,N_2$ satisfy $N_1+N_2 = N+1$. If $\m{y}\in S_0$, it can easily be shown that
\equ{\cH\m{y} = \m{A}_1 \m{S} \m{A}_2^T \label{eq:HxVD}}
and thus $\rank\sbra{\cH\m{y}}\leq K$, where $\m{S}=\diag\sbra{s_1,\dots,s_k}$ and for $j=1,2$, $\m{A}_j$ is an $N_j\times K$ Vandermonde matrix with $\mbra{\m{A}_j}_{n,k} = z_k^{n-1}$. Let
\equ{S_1 = \lbra{\m{y}\in\bC^{N}:\; \rank\sbra{\cH\m{y}} \leq K}.}
Evidently, $S_0\subseteq S_1$. By changing $S_0$ to $S_1$, the original problem \eqref{eq:genprb} is then relaxed to the following low rank Hankel optimization model:
\equ{\min_{\m{y}} \text{Loss}\sbra{\m{y},\tilde{\m{y}}}, \st \m{y}\in S_1. \label{eq:Hankel model}}
The model in \eqref{eq:Hankel model} and its variants form the Hankel-based methods.

Though the low rank Hankel model \eqref{eq:Hankel model} is a relaxation of the original problem \eqref{eq:genprb}, it has several merits. For example, EMaC, a convex relaxation of \eqref{eq:Hankel model}, has higher resolution than atomic norm methods that use the PSD-Toeplitz model \cite{chen2014robust}. Unlike the PSD-Toeplitz model, several nonconvex optimization algorithms have been proposed for spectral compressed sensing equipped with the Hankel model\cite{andersson2014new,cai2016robust,cai2019fast}. A possible reason is that the Hankel model contains fewer variables and thus is easier to initialize, which is critical for nonconvex algorithms.


\section{New Low Rank Optimization Model}  \label{sec:FB-Hankel}

\subsection{Low Rank Double Hankel Model}
In this paper, we propose to approximate the set $S_0$ in \eqref{eq:S0} by
\equ{S_2 = \lbra{\m{y}\in\bC^{N}:\; \rank\sbra{\mbra{\cH\m{y}\; | \; \m{J}_1\overline{\cH\m{y}}\m{J}_2}} \leq K}, \label{eq:S2}}
where for $j=1,2$, $\m{J}_j$ is an $N_j\times N_j$ reversal matrix with ones on the anti-diagonal and zeros elsewhere. It is seen that $\m{J}_1\overline{\cH\m{y}}\m{J}_2$ remains to be a Hankel matrix. The resulting optimization model
\equ{\min_{\m{y}} \text{Loss}\sbra{\m{y},\tilde{\m{y}}}, \st \m{y}\in S_2 \label{eq:DHankel model}}
is referred to as low rank double Hankel model.

Suppose $\m{y}\in S_0$ and recall \eqref{eq:HxVD}. Then we have
\equ{\m{J}_1\overline{\cH\m{y}}\m{J}_2 = \m{J}_1\overline{\m{A}_1 \m{S} \m{A}_2^T}\m{J}_2. \label{eq:JconjHJ}}
To proceed, it is easy to show the following identity
\equ{\m{J}_j\overline{\m{A}_j} = \m{A}_j\m{Z}^{1-N_j},\quad j=1,2, \label{eq:conjA}}
where $\m{Z} = \diag\sbra{z_1,\dots,z_K}$. Substituting \eqref{eq:conjA} into \eqref{eq:JconjHJ}, we obtain
\equ{\m{J}_1\overline{\cH\m{y}}\m{J}_2 = \m{A}_1 \overline{\m{S}} \m{Z}^{1-N}\m{A}_2^T, }
which is in the same form as \eqref{eq:HxVD}. Consequently,
\equ{\begin{split}\mbra{\cH\m{y}\; | \; \m{J}_1\overline{\cH\m{y}}\m{J}_2}
&= \mbra{\m{A}_1\m{S}\m{A}_2^T \,|\, \m{A}_1 \overline{\m{S}}\m{Z}^{1-N} \m{A}_2^T } \\
&= \m{A}_1\m{S} \begin{bmatrix}\m{A}_2 \\ \m{A}_2\m{Z}^{1-N}\tilde{\m{S}} \end{bmatrix}^T \end{split} \label{eq:dHankeldec}}
has rank no greater than $K$, where $\tilde{\m{S}} =\diag^{-2}\sbra{\text{sgn}\sbra{s_1},\dots,\text{sgn}\sbra{s_K}}$ is a unitary diagonal matrix and $\text{sgn}\sbra{s} = \frac{s}{\abs{s}}$ for $s\neq 0$. We therefore conclude
\equ{S_0 \subseteq S_2\subseteq S_1}
which implies that, as compared to the Hankel model \eqref{eq:Hankel model}, the proposed double Hankel model \eqref{eq:DHankel model} is a tighter relaxation of the original problem \eqref{eq:genprb}. More differences between them will be shown in the ensuing subsection.

It is worth noting that the proposed double Hankel model is partially inspired by the forward-backward processing technique \cite{evans1982application,williams1988improved}. In particular, let $\breve{\m{y}}\in\bC^N$ be such that
\equ{\m{J}_1\overline{\cH\m{y}}\m{J}_2 = \cH\breve{\m{y}}. \label{eq:Htildex}}
It is seen that $\breve{\m{y}}$ is the conjugated backward version of $\m{y}$ satisfying
\equ{\breve y_j = \overline{y}_{N-j+1}=\sum_{k=1}^K \overline s_k \overline z_k^{N-j} = \sum_{k=1}^K \overline s_k z_k^{j-N} \label{eq:conjback}, \quad j=1,\dots,N}
by making use of \eqref{eq:S0}.
Equation \eqref{eq:conjback} implies that the virtual signal $\breve{\m{y}}$ is composed of the same spectral poles $\lbra{z_k}$ as $\m{y}$, which is the key observation underlying the forward-backward processing technique.

\subsection{Properties} \label{sec:properties}
In this subsection we analyze the low rank Hankel and double Hankel models and illustrate advantages of the latter. To this end, we let
\equ{S'_1 = \lbra{\m{y}\in\bC^{N}:\; y_j = \sum_{k=1}^K s_k z_k^{j-1},\; s_k\in\bC,\;0\neq z_k\in\bC} \label{eq:S1p}}
that is obtained by removing the constraint $\abs{z_k}=1$ in $S_0$ and thus $S_0\subset S'_1$. To link $S'_1$ to $S_1$, on one hand, we have \eqref{eq:HxVD} for any $\m{y}\in S'_1$ and thus $S'_1\subseteq S_1$. On the other hand, it follows from the Kronecker's theorem \cite{rochberg1987toeplitz,ellis1992factorization} that if $\m{y}\in S_1$ and $K < \min\lbra{N_1,\;N_2}$, then $\m{y}\in S_1'$ except for degenerate cases. Therefore, $S_1$ and $S_1'$ are approximately identical given $K < \min\lbra{N_1,\;N_2}$. This implies that {\em the Hankel model \eqref{eq:Hankel model} can be viewed as a relaxation of \eqref{eq:genprb} by removing the constraint $\abs{z_k}=1$}. In other words, by using the Hankel model \eqref{eq:Hankel model}, we have actually abandoned the prior knowledge that the spectral poles $\lbra{z_k}$ lie on the unit circle. This claim is consistent with the observation in \cite{hua1990matrix,chen2014robust,andersson2014new} that the matrix pencil method and the Hankel model also apply to damped spectrally sparse signals for which the magnitudes of the spectral poles $\lbra{\abs{z_k}}$ are unknown.

We next study the double Hankel model in \eqref{eq:DHankel model}. Assume $\m{y}\in S_2$ and $K< \min\lbra{N_1, N_2}$ and recall \eqref{eq:Htildex} and \eqref{eq:conjback}. Then we have
\equ{\rank\sbra{\mbra{\cH\m{y}\; | \; \cH\breve{\m{y}}}} = \rank\sbra{\mbra{\cH\m{y}\; | \; \m{J}_1 \overline{\cH\m{y}}\m{J}_2}}\leq K. \label{eq:rankXe}}
It follows that $\rank\sbra{\cH\m{y}}\leq K< \min\lbra{N_1, N_2}$ and $\rank\sbra{\cH\breve{\m{y}}}\leq K< \min\lbra{N_1, N_2}$. Applying the Kronecker's theorem results in $\m{y}, \breve{\m{y}} \in S_1'$ almost surely, to be specific, there exist $\lbra{s_k,z_k\in \bC}_{k=1}^K$ and $\lbra{s_k',z_k'\in \bC}_{k=1}^K$ with $z_k,z_k'\neq 0$ such that for $j=1,\dots,N$,
{\lentwo\equa{ y_j
&=& \sum_{k=1}^K s_k {z}_k^{j-1}, \label{eq:xj}\\\overline y_{N-j+1}=\breve{y}_j
&=& \sum_{k=1}^K s'_k {z}_k'^{j-1}, \label{eq:cxmj}
}}yielding that
\equ{y_j= \sum_{k=1}^K s_k {z}_k^{j-1} = \sum_{k=1}^K \overline{s'_k{z}_k'^{N-1}} {z}_k'^{\star j-1}, \label{eq:xjident}}
where we denote $z^{\star} = \overline z^{-1}$ for $z\in\bC$. Equation \eqref{eq:xjident} says that $\m{y}$ admits two different decompositions with each consisting of $K$ poles. Recall $N=N_1+N_2-1\geq 2K+1$, and thus any $N\times 2K$ Vandermonde matrix with distinct poles has full column rank. Using this fact, we conclude that the two decompositions in \eqref{eq:xjident} must be identical, implying without loss of generality that for some $K_0\leq K$,
{\lentwo\equa{z_k
&=& {z}_k'^{\star},\quad s_k=\overline{s'_kz_k'^{N-1}},\quad k=1,\dots, K_0, \notag \\ s_k
&=& s'_k = 0, \quad k=K_0+1,\dots, K. \label{eq:map}}

Using \eqref{eq:xj}, \eqref{eq:cxmj}, \eqref{eq:Hankel} and \eqref{eq:map}, we obtain
\equ{\begin{split}
&\mbra{\cH\m{y}\; | \; \cH\breve{\m{y}}} \\
&= \mbra{\m{A}\sbra{\m{z}}\m{S}\m{A}^T\sbra{\m{z}}\; | \; \m{A}\sbra{\m{z}'}\m{S}'\m{A}^T\sbra{\m{z}'}}\\
&= \mbra{\m{A}\sbra{\m{z}}\m{S}\m{A}^T\sbra{\m{z}}\; | \; \m{A}\sbra{\m{z}^{\star}}\m{S}'\m{A}^T\sbra{\m{z}^{\star}}}\\
&= \mbra{\m{A}\sbra{\m{z}}\; | \; \m{A}\sbra{\m{z}^{\star}}} \begin{bmatrix} \m{S}\m{A}^T\sbra{\m{z}} & \\ & \m{S}'\m{A}^T\sbra{\m{z}^{\star}} \end{bmatrix}, \end{split}}
where $\m{S}'=\diag\sbra{s'_1,\dots,s'_K}$.
Consequently, it is seen that
$\rank\sbra{\mbra{\cH\m{y}\; | \; \cH\breve{\m{y}}}}\leq K$ equals the number of distinct poles in $\lbra{z_k}_{k=1}^{K_0} \cup \lbra{z_k^{\star}}_{k=1}^{K_0}$. Using the fact that $z_k=z_k^{\star}$ if and only if $\abs{z_k}=1$, we have $\m{y}\in S'_2$, where
\equ{S'_2 = \lbra{\m{y}\in S'_1 \text{ where either $\abs{z_k}=1$ or $z_k,z_k^{\star}$ appear in pair}}. \label{eq:S2p}}

It can be shown by arguments similar to those above that the converse, $S'_2 \subseteq S_2$, is also true. Therefore, $S'_2$ and $S_2$ are approximately identical, and we conclude that {\em the double Hankel model \eqref{eq:DHankel model} can be viewed as a relaxation of \eqref{eq:genprb} by allowing appearance of pairs of poles $\lbra{z_k, z_k^{\star}}$ if they are not on the unit circle}.

Note that if we write $z_k=r_k e^{i2\pi f_k}$ in the polar coordinate, then $z_k^{\star} = r_k^{-1} e^{i2\pi f_k}$ and thus the pair of estimated poles $\lbra{z_k, z_k^{\star}}$ share an identical frequency, implying that they correspond to a same spectral component given the prior knowledge that the true spectral poles lie on the unit circle. Intuitively, such pair of estimated poles is obtained only when two spectral frequencies are closely located so that they cannot be separated. If all the frequencies are properly separated, it is expected that all of the estimated poles obtained by using the double Hankel model lie on the unit circle, which, as we will see, is consistent with our numerical results presented in Section \ref{sec:simulation}.

In the case $K=1$, we have the following result.
\begin{prop} If $K=1$ and $N_1,N_2>1$, then
\equ{S_0=S'_2=S_2\subset S'_1\subset S_1.} \label{prop:K1}
\end{prop}
\begin{proof} First, when $K=1$ it is evident that $S_0=S'_2\subset S'_1$. We show next
\equ{S_1 = S'_1\cup \lbra{\begin{bmatrix} y \\ 0 \\ \vdots \\ 0 \end{bmatrix}:\; y\in\bC} \cup \lbra{\begin{bmatrix} 0 \\ \vdots \\ 0 \\ y \end{bmatrix}:\; y\in\bC}\supset S'_1. \label{eq:S1}}
To do so, we assume without loss of generality that $N_1=N_2$ and $\cH\m{y}$ is a square matrix (Otherwise, we can consider the leading square submatrix of $\cH\m{y}$ of size $\min\sbra{N_1,N_2}$ and draw the same conclusion.) Let $\cH_1$ be the Hankel matrix obtained by removing the last row and column of $\cH\m{y}$. If $\rank\sbra{\cH_1}=0$, which implies $\cH_1=\m{0}$, we must have that all but the last entry of $\m{y}$ are zero. If $\rank\sbra{\cH_1}=\rank\sbra{\cH\m{y}}=1$, applying \cite[Theorem 3.1]{ellis1992factorization}, we have $\m{y}\in S'_1$, or all but the first entry of $\m{y}$ are zero.

Now it suffices to show $S_2=S'_2$, or $S_2\subseteq S'_2$, since it is evident that $S'_2\subseteq S_2$. For any $\m{y}\in S_2$, it is evident that $\m{y}\in S_1$. Then by \eqref{eq:S1} we must have $\m{y}\in S'_1$ since otherwise $\rank\sbra{\mbra{\cH\m{y}\,|\, \m{J}_1\overline{\cH\m{y}}\m{J}_2}}=2$ that implies that $\m{y}\notin S_2$. Consequently, by the same arguments as those in Subsection \ref{sec:properties}, we have $\m{y}\in S'_2$, completing the proof.
\end{proof}

It is implied by Proposition \ref{prop:K1} that when $K=1$ the double Hankel model is equivalent to the original problem \eqref{eq:genprb} but the Hankel model is not.

\subsection{Comparison With Previous Models}
For the proposed double Hankel model and the previous PSD-Toeplitz and Hankel models, we present in Table \ref{tab:model} the number of (real-valued) optimization variables in the signal space [see \eqref{eq:toeplitz_model}, \eqref{eq:Hankel model} and \eqref{eq:DHankel model}] and in the parameter space [see \eqref{eq:S0}, \eqref{eq:S1p} and \eqref{eq:S2p}] respectively. The former quantity measures complexity of an optimization model, and the latter reflects the model accuracy.

The PSD-Toeplitz model is equivalent to the original problem \eqref{eq:genprb} and thus includes $3K$ optimization variables in the parameter space. To achieve such equivalence, $2N$ new variables in the Hermitian Toeplitz matrix are introduced, resulting in $4N$ variables in total. The Hankel model contains a minimum number $2N$ of variables in $\m{y}$, while its solution space is enlarged and $K$ new variables (the amplitudes of poles $\lbra{\abs{z_k}}_{k=1}^K$) are introduced in the parameter space. In contrast to those above, the double Hankel model possesses the same variables as the Hankel model in the signal space and the same number of variables as the PSD-Toeplitz model in the parameter space.

\begin{table*}
\centering
\caption{Number of real-valued optimization variables of three optimization models in signal and parameter spaces. The former quantity measures model complexity (the less the better), and the latter reflects model accuracy (the ground truth is $3K$).}
\begin{tabular}{|c|c|c|}
\hline
\textbf{Optimization models}& \textbf{\#variables in signal space} & \textbf{\#variables in parameter space}\\ 
\hline
\rule{0pt}{15pt}  PSD-Toeplitz & $4N$ & $3K$ \\
\hline
\rule{0pt}{15pt}  Hankel & $2N$ & $4K$ \\ \hline
\rule{0pt}{15pt} \textbf{Double Hankel (proposed)}& $2N$ & $3K$ \\
\hline
\end{tabular}
\label{tab:model}
\end{table*}

\section{Convex Relaxation Approaches} \label{sec:demac}
In this section we exploit the double Hankel model \eqref{eq:DHankel model} and propose convex relaxation approaches for spectral compressed sensing with provable accuracy. It is worth noting that we do not assume that the number of frequencies $K$ is known. Instead, certain prior knowledge on the measurement noise will be assumed. Our approaches are mainly inspired by \cite{chen2014robust} where the low rank Hankel model is considered, and they are named as doubly enhanced matrix completion (DEMaC) following EMaC in \cite{chen2014robust}.

\subsection{DEMaC in Absence of Noise}
We first consider the noiseless case where we have access to $\cP_{\Omega}\sbra{\m{y}^o}$,
where $\Omega\subseteq\lbra{1,\dots,N}$ is the sampling index set of size $M\leq N$, and $\cP_{\Omega}$ is the projection onto the subspace supported on $\Omega$ that sets all entries of $\m{y}$ out of $\Omega$ to zero. By swapping the objective and the constraint, the double Hankel model \eqref{eq:DHankel model} is equivalently written as:
\equ{\min_{\m{y}} \rank\sbra{\mbra{\cH\m{y}\,|\, \m{J}_1\overline{\cH\m{y}}\m{J}_2}}, \st \cP_{\Omega}\sbra{\m{y}-\m{y}^o} = \m{0}. \label{eq:dHankel2}}

Following the literature on low rank matrix recovery, we relax the rank function in \eqref{eq:dHankel2} to the nuclear norm and obtain
\equ{\begin{split}\text{(DEMaC)}\quad &\min_{\m{y}} \norm{\mbra{\cH\m{y}\,|\, \m{J}_1\overline{\cH\m{y}}\m{J}_2}}_{\star}\\
&\st \cP_{\Omega}\sbra{\m{y}-\m{y}^o} = \m{0}. \end{split} \label{eq:demac}}
The DEMaC problem in \eqref{eq:demac} is convex and can be written as semidefinite programming (SDP) since the nuclear norm $\norm{\m{X}}_{\star}$ of matrix $\m{X}$ can be cast as the SDP
\equ{\min_{\m{P},\m{Q}} \frac{1}{2}\tr\sbra{\m{P}} + \frac{1}{2}\tr\sbra{\m{Q}}, \st \begin{bmatrix} \m{P} & \m{X}^H \\ \m{X} & \m{Q}\end{bmatrix}\geq \m{0}. \label{eq:nuclearnormSDP}}
where $\m{P}$ and $\m{Q}$ are Hermitian matrices.
Therefore, DEMaC can be solved using off-the-shelf SDP solvers such as SDPT3 \cite{toh1999sdpt3}.

\subsection{Theoretical Guarantee With Full Sampling}
Assume $\Omega = \lbra{1,\dots,N}$ and there is no noise. Then we have full knowledge of the true signal $\m{y}^o$ and thus it is trivial to solve DEMaC in \eqref{eq:demac}. We next study the condition for exact recovery of the distinct frequencies $\lbra{f_k}$. It is worth noting that an equivalent problem has been studied in \cite{pillai1989forward,choi2002conditions} in the context of forward-backward spatial smoothing for coherent direction-of-arrival (DOA) estimation, where the frequencies are estimated from a covariance matrix that equals
\equ{\begin{split}
&\mbra{\cH\m{y}^o\,|\, \m{J}_1\overline{\cH\m{y}^o}\m{J}_2}\mbra{\cH\m{y}^o\,|\, \m{J}_1\overline{\cH\m{y}^o}\m{J}_2}^H \\
&= \cH\m{y}^o\cdot \sbra{\cH\m{y}^o}^H + \m{J}_1\overline{\cH\m{y}^o\cdot \sbra{\cH\m{y}^o}^H}\m{J}_1 \end{split}}
using a subspace method such as ESPRIT.

Let us recall \eqref{eq:dHankeldec} and assume $K \leq N_1-1$. We compute the truncated singular value decomposition (SVD) of the double Hankel matrix as
\equ{\mbra{\cH\m{y}^o\,|\, \m{J}_1\overline{\cH\m{y}^o}\m{J}_2} = \m{U}\m{\Lambda}\m{V}^H. \label{eq:SVD}}
If the double Hankel matrix has the maximal rank $K$, then $\m{A}_1$ in \eqref{eq:dHankeldec} and $\m{U}$ share an identical range space that is known as the signal subspace from which the frequencies can be uniquely computed using ESPRIT. Now our question becomes under what conditions the double Hankel matrix or $\begin{bmatrix}\m{A}_2 \\ \m{A}_2\m{Z}^{1-N}\tilde{\m{S}} \end{bmatrix}$ in \eqref{eq:dHankeldec} has rank $K$.

Evidently, if $K \leq N_2$, then $\m{A}_2$ has full column rank and thus $\rank\sbra{\begin{bmatrix}\m{A}_2 \\ \m{A}_2\m{Z}^{1-N}\tilde{\m{S}} \end{bmatrix}}=K$.

If $N_2< K \leq 2N_2$ and for arbitrary distinct $\lbra{f_k}$, it follows from \cite{choi2002conditions} that $\begin{bmatrix}\m{A}_2 \\ \m{A}_2\m{Z}^{1-N}\tilde{\m{S}} \end{bmatrix}$ is rank-deficient only if the phases $\lbra{\text{sgn}\sbra{s_k}}$ of the spectral components satisfy certain equations that will be violated with probability one if $\lbra{\text{sgn}\sbra{s_k}}$ are jointly distributed according to some absolutely continuous distribution.

To sum up, we conclude the following result.

\begin{thm} Assume $\Omega = \lbra{1,\dots,N}$ and no noise is present. Then DEMaC is able to exactly recover the distinct frequencies $\lbra{f_k}$ if
\equ{K\leq \min\sbra{N_1-1,\;N_2}\leq \lfloor \frac{N}{2}\rfloor, \label{eq:K1}}
or with probability one if
\equ{K\leq \min\sbra{N_1-1,\;2N_2}\leq \lfloor \frac{2N}{3}\rfloor \label{eq:K2}}
and if the phases $\lbra{\text{sgn}\sbra{s_k}}$ of the spectral components are jointly distributed according to some absolutely continuous distribution on the $K$-dimensional torus.
\label{Thm1}
\end{thm}

Note that all equalities in \eqref{eq:K1}--\eqref{eq:K2} can be achieved, implying that up to $\lfloor \frac{N}{2}\rfloor$ frequencies can be recovered deterministically and up to $\lfloor \frac{2N}{3}\rfloor$ frequencies almost surely. In contrast to this, no more than $\lfloor \frac{N}{2}\rfloor$ frequencies can be recovered if the Hankel model is used, as in \cite{chen2014robust}. Moreover, to make the number of recoverable frequencies as large as possible, we should choose $N_1$, $N_2$ such that $2N_2\geq N_1\geq N_2$ according to \eqref{eq:K1}--\eqref{eq:K2}, implying
\equ{\frac{2N}{3}\geq N_1\geq \frac{N}{2}. \label{eq:N1choice}}
This choice also makes sense in the compressive sampling settings as shown later.

The SVD plus ESPRIT method mentioned above can also be used in general settings that we will study next to retrieve the estimated poles $\lbra{\hat z_k}$ from the solution of the double Hankel matrix once DEMaC is numerically solved.

\subsection{Theoretical Guarantee With Compressive Sampling}
When partial entries of $\m{y}^o$ are observed in absence of noise, we show in this subsection that DEMaC has similar theoretical guarantee as EMaC. We first introduce some notations similar to those in \cite{chen2014robust}. Recall \eqref{eq:dHankeldec}, and let
\equ{\tilde{\m{A}}_2 = \begin{bmatrix}\m{A}_2 \\ \m{A}_2\m{Z}^{1-N}\tilde{\m{S}} \end{bmatrix} \label{eq:A2tilde}}
and
\equ{\m{G}_1 = \frac{1}{N_1} \m{A}_1^H\m{A}_1,\quad \m{G}_2 = \frac{1}{2N_2} \overline{\tilde{\m{A}}_2^H\tilde{\m{A}}_2}. \label{eq:G1G2}}
The double Hankel matrix $\mbra{\cH\m{y}\,|\, \m{J}_1\overline{\cH\m{y}}\m{J}_2}$ is said to obey the incoherence property with parameter $\mu_1$ if
\equ{\lambda_{\text{min}}\sbra{\m{G}_1}\geq \frac{1}{\mu_1} \text{ and } \lambda_{\text{min}}\sbra{\m{G}_2}\geq \frac{1}{\mu_1}, \label{eq:incoh}}
where $\lambda_{\text{min}}\sbra{\cdot}$ denotes the minimum eigenvalue of a matrix. Note that the incoherence property is defined in \cite{chen2014robust} based on $\m{G}_1$ above and
\equ{\m{G}'_2 = \frac{1}{N_2}\overline{\m{A}_2^H\m{A}_2}.}
It is seen from \eqref{eq:A2tilde} and \eqref{eq:G1G2} that
\equ{\begin{split}\m{G}_2
&= \frac{1}{2N_2}\cdot \overline{\m{A}_2^H\m{A}_2 + \overline{\tilde{\m{S}}}\m{Z}^{N-1} \m{A}_2^H\m{A}_2\m{Z}^{1-N} \tilde{\m{S}} }\\
&= \frac{1}{2}\mbra{\m{G}'_2 +\tilde{\m{S}}\m{Z}^{1-N} \m{G}'_2 \m{Z}^{N-1} \overline{\tilde{\m{S}}} }, \end{split} \label{eq:G2}}
and thus
\equ{\begin{split}\lambda_{\text{min}}\sbra{\m{G}_2}
&\geq \frac{1}{2} \mbra{ \lambda_{\text{min}}\sbra{\m{G}'_2} + \lambda_{\text{min}}\sbra{\tilde{\m{S}}\m{Z}^{1-N} \m{G}'_2 \m{Z}^{N-1} \overline{\tilde{\m{S}}}} } \\
&= \frac{1}{2} \mbra{ \lambda_{\text{min}}\sbra{\m{G}'_2} + \lambda_{\text{min}}\sbra{\m{G}'_2} }\\
&= \lambda_{\text{min}}\sbra{\m{G}'_2}, \end{split}}
where the first equality holds since both $\m{Z}$ and $\tilde{\m{S}}$ are unitary matrices. Consequently, the incoherence property defined here with respect to DEMaC is easier to satisfy than that with EMaC in \cite{chen2014robust}.


Let
\equ{c_s = \max\lbra{\frac{N}{N_1},\; \frac{N}{2N_2}} \label{eq:cs}}
that measures how close the double Hankel matrix is to a square matrix. The factor 2 in the second denominator in \eqref{eq:cs} comes along with the definition of $\m{G}_2$ in \eqref{eq:G1G2}.

We have the following result that is in parallel with \cite[Theorem 1]{chen2014robust} and shows that, like EMaC, DEMaC perfectly recovers the spectrally sparse signal from $O\sbra{K\log^4N}$ random samples.

\begin{thm} Let $\m{y}^o$ be given in \eqref{eq:model}, and $\Omega$ the random index set of size $M$. Suppose that the incoherence property \eqref{eq:incoh} holds and that all samples are noiseless. Then there exists a universal constant $c_1>0$ such that $\m{y}^o$ is the unique solution to DEMaC in \eqref{eq:demac} with probability exceeding $1-N^{-2}$, provided that
\equ{M > c_1\mu_1c_sK \log^4 N. \label{eq:bound}} \label{thm:noiseless}
\end{thm}

The proof of Theorem \ref{thm:noiseless} is similar to that of \cite[Theorem 1]{chen2014robust}, and the latter is inspired by \cite{gross2011recovering} in which the general low rank matrix recovery problem is studied via affine nuclear norm minimization. In particular, as in \cite{gross2011recovering,chen2014robust}, DEMaC is first rewritten as a nuclear norm minimization problem under affine (and additional) constraints that are introduced to identify the sampling basis and capture the Hankel structures of the double Hankel matrix to recover. Then, it is shown that the sampling basis fulfils the incoherence condition with respect to the tangent space of the true low rank matrix. Finally, a dual certificate is constructed and rigorously verified via the golfing scheme \cite{gross2011recovering}. The new challenges in our proof are to identify the sampling basis with the double Hankel model and to properly deal with the complex conjugate operator in the double Hankel matrix that leads to non-affine constraints. The detailed proof is complicated and deferred to the Appendix.

The sample size shown in Theorem \ref{thm:noiseless} is an increasing function of $c_s$ defined in \eqref{eq:cs}. Consequently, we should choose $N_1$, $N_2$ such that $c_s$ is small. Therefore, the choice in \eqref{eq:N1choice} is also appropriate in this compressive sampling setting.

\subsection{DEMaC With Bounded Noise}
We now consider the noisy case where the acquired samples $\cP_{\Omega}\sbra{\tilde{\m{y}}}$ are given by \eqref{eq:model}, with $\twon{\cP_{\Omega}\sbra{\m{e}}}\leq \eta$. Consequently, we solve the following noisy version of DEMaC:
\equ{\begin{split}\text{(Noisy-DEMaC)}\quad &\min_{\m{y}} \norm{\mbra{\cH\m{y}\,|\, \m{J}_1\overline{\cH\m{y}}\m{J}_2}}_{\star}\\
&\st \norm{\cP_{\Omega}\sbra{\m{y}-\tilde{\m{y}}}} \leq \eta. \end{split} \label{eq:demacnoise}}

The following result is a consequence of combining Theorem \ref{thm:noiseless} and \cite[Theorem 2]{chen2014robust}. Its proof can be derived by slightly modifying the proof of \cite[Theorem 2]{chen2014robust} based on technical results in the proof of Theorem \ref{thm:noiseless} and thus is omitted.

\begin{thm} Suppose $\tilde{\m{y}}$ is a noisy copy of $\m{y}^o$ that satisfies $\norm{\cP_{\Omega}\sbra{\m{y}^o-\tilde{\m{y}}}} \leq \eta$. Under the assumptions of Theorem \ref{thm:noiseless}, the solution $\hat{\m{y}}$ to Noisy-DEMaC in \eqref{eq:demacnoise} satisfies
\equ{\frobn{\cH\hat{\m{y}} - \cH \m{y}^o} \leq 5N^3\eta}
with probability exceeding $1-N^{-2}$. \label{thm:noisy}
\end{thm}

\subsection{DEMaC With Sparse Noise}
It is an important research topic to deal with outliers in data samples that can be caused for example by saturation in data quantization and abnormal sensor behaviors. We assume in this case that a constant portion of the acquired samples $\cP_{\Omega}\sbra{\tilde{\m{y}}}$ are outliers that are modelled via additive sparse noise. As in \cite{chen2014robust}, we modify DEMaC to include the sparse noise by solving the following problem:
\equ{\begin{split}\text{(Robust-DEMaC)}\quad &\min_{\m{y}, \m{e}} \norm{\mbra{\cH\m{y}\,|\, \m{J}_1\overline{\cH\m{y}}\m{J}_2}}_{\star} + 2\lambda\norm{\cH\m{e}}_{1}\\
&\st \cP_{\Omega}\sbra{\m{y}+\m{e}} = \cP_{\Omega}\sbra{\tilde{\m{y}}}. \end{split} \label{eq:demacrobust}}
Note in \eqref{eq:demacrobust} that $\norm{\cH\m{e}}_{1}=\onen{\text{vec}\sbra{\cH\m{e}}}$ is defined as the elementwise $\ell_1$ norm and thus $2\norm{\cH\m{e}}_{1} = \norm{\mbra{\cH\m{e}\,|\, \m{J}_1\overline{\cH\m{e}}\m{J}_2}}_{1}$.

We have the following result by simply combining Theorem \ref{thm:noiseless} and \cite[Theorem 3]{chen2014robust} whose proof is omitted.
\begin{thm} Suppose $\tilde{\m{y}}$ is a noisy copy of $\m{y}^o$. Let $\Omega$ be the random index set of size $M$, and conditioning on $\Omega$ each sample in $\Omega$ is corrupted by noise independently with conditional probability $\tau\leq 0.1$. Set $\lambda = \frac{1}{\sqrt{M\log N}}$. Then there exists a numerical constant $c_1>0$ depending only on $\tau$ such that if $\eqref{eq:incoh}$ holds and
\equ{M > c_1\mu_1^2c_s^2K^2 \log^3 N,}
then $\m{y}^o$ is the unique solution to Robust-DEMaC in \eqref{eq:demacrobust} with probability exceeding $1-N^{-2}$.
\label{thm:robust}
\end{thm}

\section{Extension to Dimension Two and Above}
In this section, we extend the low rank double Hankel model and the DEMaC approaches from the 1-D to arbitrary-dimensional spectral compressed sensing. Without loss of generality and for simplicity of notations, we first consider the 2-D case. Certain notations such as $N_1$, $N_2$ will be redefined. Corresponding to \eqref{eq:model}, a noiseless 2-D spectrally sparse signal is given by a 2-way array (or matrix) $\m{y}$ whose entries are
\equ{y_{j_1,j_2} = \sum^{K}_{k=1}s_{k} z^{j_1-1}_{k,1} z^{j_2-1}_{k,2},  \quad  \abs{z_{k,1}}=\abs{z_{k,2}}=1,}
where $1\leq j_1\leq N_1$, $1\leq j_2\leq N_2$, and the total sample size is $N=N_1 \cdot N_2$.

Let $\m{y}_{j_1,:}$ denote the $N_2\times 1$ vector composed of $y_{j_1,1}, \dots, y_{j_1,N_2}$ for fixed $j_1$. As in \eqref{eq:Hankel}, $\cH\m{y}_{j_1,:}$ denotes an $N_{2,1}\times N_{2,2}$ Hankel matrix formed by using $\m{y}_{j_1,:}$ where $N_{2,1}+N_{2,2}=N_2+1$. Then we define a 2-level Hankel (or Hankel-block-Hankel) matrix $\cH \m{y}$ as an $N_{1,1}\times N_{1,2}$ block Hankel matrix formed by $\cH\m{y}_{1,:},\dots, \cH\m{y}_{N_1,:}$ where $N_{1,1}+N_{1,2}=N_1+1$, to be specific,
\equ{\cH\m{y}= \begin{bmatrix} \cH\m{y}_{1,:} & \cH\m{y}_{2,:} & \dots & \cH\m{y}_{N_{1,2},:} \\ \cH\m{y}_{2,:} & \cH\m{y}_{3,:} & \dots & \cH\m{y}_{N_{1,2}+1,:} \\ \vdots & \vdots & \ddots & \vdots \\ \cH\m{y}_{N_{1,1},:} & \cH\m{y}_{N_{1,1}+1,:} & \dots & \cH\m{y}_{N_{1},:}\end{bmatrix}. \label{eq:Hankel2}}

For $j=1,2$ and $l=1,2$, we define $\m{A}_{jl}$ as an $N_{jl}\times K$ Vandermonde matrix with $\mbra{\m{A}_{jl}}_{n,k}=z_{k,j}^{n-1}$ and let $\m{A}_l = \m{A}_{1l}\star \m{A}_{2l}$, where $\star$ denotes the Khatri-Rao (or column-wise Kronecker) product. Consequently, the $k$th column of $\m{A}_j$ represents a sampled 2-D sinusoid with pole $\m{z}_k=\sbra{z_{k,1}, z_{k,2}}$. Using these notations, it can be readily verified that (we omit the details)
\equ{\cH\m{y} = \m{A}_1 \m{S} \m{A}_2^T, \label{eq:HxVD2}}
which is identical to \eqref{eq:HxVD}. This implies that the 2-level Hankel matrix $\cH\m{y}$ is low rank.

For $l=1,2$, let $\m{J}_l$ be the $N_{1l}N_{2l}\times N_{1l}N_{2l}$ reversal matrix that can be written as a Kronecker product, $\m{J}_l = \m{J}_{1l}\otimes\m{J}_{2l}$, where $\m{J}_{jl}$ is the $N_{jl}\times N_{jl}$ reversal matrix for $j=1,2$ and $l=1,2$. By applying the following identity for matrices $\m{A},\m{B}, \m{C}, \m{D}$ of proper dimension
\equ{\sbra{\m{A}\otimes\m{B}} \sbra{\m{C}\star\m{D}} = \sbra{\m{A}\m{C}} \star \sbra{\m{B}\m{D}}, }
we have that for $l=1,2$,
\equ{\begin{split}\m{J}_l \overline{\m{A}_l}
&= \sbra{\m{J}_{1l}\otimes\m{J}_{2l}} \sbra{ \overline{\m{A}_{1l}}\star \overline{\m{A}_{2l}}}\\
&= \sbra{\m{J}_{1l}\overline{\m{A}_{1l}}} \star \sbra{\m{J}_{2l}\overline{\m{A}_{2l}}}\\
&= \sbra{\m{A}_{1l}\m{Z}_1^{1-N_{1l}}} \star \sbra{\m{A}_{2l}\m{Z}_2^{1-N_{2l}}}\\
&= \sbra{\m{A}_{1l} \star \m{A}_{2l}} \m{Z}_1^{1-N_{1l}}\m{Z}_2^{1-N_{2l}} \\
&= \m{A}_l \m{Z}_1^{1-N_{1l}}\m{Z}_2^{1-N_{2l}}, \end{split} \label{eq:JA2}}
where $\m{Z}_j = \diag\sbra{z_{1j},\dots,z_{kj}}$ for $j=1,2$, the third equality follows from \eqref{eq:conjA}, and the fourth equality holds since $\m{Z}_1$, $\m{Z}_2$ are diagonal matrices. It immediately follows from \eqref{eq:HxVD2} and \eqref{eq:JA2} that
\equ{\begin{split}\m{J}_1 \overline{\cH \m{y}} \m{J}_2
&= \m{J}_1 \overline{\m{A}_1\m{S}\m{A}_2^T} \m{J}_2 \\
&= \m{J}_1 \overline{\m{A}_1} \cdot \overline{\m{S}} \sbra{\m{J}_2 \overline{\m{A}_2}}^T\\
&= \m{A}_1 \m{Z}_1^{1-N_{11}}\m{Z}_2^{1-N_{21}} \overline{\m{S}} \m{Z}_2^{1-N_{22}}\m{Z}_1^{1-N_{12}}\m{A}_2^T \\
&= \m{A}_1 \overline{\m{S}} \m{Z}_1^{1-N_{1}}\m{Z}_2^{1-N_{2}}\m{A}_2^T, \end{split} \label{eq:JHyJ2}}
where the last equality holds since $\m{Z}_1$, $\m{Z}_2$, $\m{S}$ are all diagonal matrices. Therefore, as in the 1-D case, the new 2-level Hankel matrix $\m{J}_1 \overline{\cH \m{y}} \m{J}_2 $ admits a similar decomposition as $\cH \m{y}$.

As in the 1-D case, by replacing the 2-level Hankel matrix $\cH \m{y}$ by the double 2-level Hankel matrix $\mbra{\cH\m{y}\,|\, \m{J}_1\overline{\cH\m{y}}\m{J}_2}$, we are able to propose the double Hankel model and the DEMaC approaches for 2-D spectral compressed sensing. They generalize the Hankel model and EMaC in \cite{andersson2014new,cai2019fast,chen2014robust}. Note that the theoretical results for DEMaC also hold in the 2-D case as in \cite{chen2014robust}.

The double Hankel model and the DEMaC approach can be generalized to arbitrary dimension. In dimension $d\geq 3$, the noiseless data $\m{y}$ are given by a $d$-way array that can be used to form a $d$-level Hankel matrix $\cH \m{y}$ as in \eqref{eq:Hankel2}, to be specific, a $d$-level Hankel matrix is a block Hankel matrix with each block being a $\sbra{d-1}$-level Hankel matrix. Then we can similarly define $\m{A}_l$, $l=1,2$ as the Khatri-Rao product of $d$ Vandermonde matrices, of which each column represents a sampled $d$-D sinusoid, and derive identities \eqref{eq:HxVD2} and \eqref{eq:JHyJ2}.

\section{Numerical Results} \label{sec:simulation}

\subsection{Model Efficiency}
In this subsection we validate efficiency of the double Hankel model proposed in Section \ref{sec:FB-Hankel} in restricting the spectral poles on the unit circle. To this end, we present an iterative hard thresholding (IHT) algorithm to solve \eqref{eq:DHankel model} that is modified from the one rooted on the Hankel model proposed in \cite{cai2019fast}. In particular, we consider the noisy, full sampling case and use the least square loss function. The double Hankel model that we need to solve is thus given by
\equ{\min_{\m{y}} \frac{1}{2}\norm{\m{y}-\tilde{\m{y}}}^2_{2}, \st \rank\sbra{\cH_{\text{D}}\sbra{\m{y}}} \leq K, \label{eq:noisyFBHankel}}
where $\cH_{\text{D}}\sbra{\m{y}}=\mbra{\cH\m{y}\; | \; \m{J}_1\overline{\cH\m{y}}\m{J}_2}$.

Our algorithm is illustrated in Algorithm \ref{alg:IHT}. It starts with the initialization  $\m{y}_0=\widetilde{\m{y}}$. Then gradient descent is used to update the signal estimate at each iteration with step size $\alpha_t = \frac{1}{\sqrt{t}}$. After the double Hankel matrix $\m{D}_t$ is formed, the hard thresholding operator $\m{\Gamma}_{K}$ is applied to obtain the best rank-$K$ approximation, $\m{G}_{t}$, of $\m{D}_t$ by setting all but its largest $K$ singular values to zero. The new estimate $\m{y}_{t+1}$ is obtained as the minimizer to the problem $\min_{\m{y}}\twon{\cH_{\text{D}}\m{y} - \m{G}_{t}}$, where $\cH_{\text{D}}^{\dag}$ denotes the pseudoinverse of the double Hankel operator. Note that as compared to \cite{cai2019fast} the Hankel operator $\cH$ is changed to the double Hankel operator $\cH_{\text{D}}$ in Algorithm \ref{alg:IHT}, and a descending step size is adopted that empirically ensures feasibility of the solution once converged.

\begin{algorithm}
\caption{The IHT algorithm with double Hankel model}
\label{alg:IHT}
\begin{algorithmic}
\REQUIRE $\m{y}_{0}=\widetilde{\m{y}}$
\FOR{$t = 1,...$}
\STATE $\m{D}_{t}=\cH_{\text{D}}\sbra{\m{y}_{t}+\alpha_t\sbra{\widetilde{\m{y}}-\m{y_t}}}$
\STATE $\m{G}_{t}=\m{\Gamma}_{K}\sbra{\m{D}_{t}}$
\STATE $\m{y}_{t+1}=\cH_{\text{D}}^{\dag}\sbra{\m{G_{t}}}$
\ENDFOR
\end{algorithmic} \label{alg:IHT}
\end{algorithm}

\begin{table*}
\centering
\fontsize{7}{8}\selectfont
\caption{Details of the three failures with the double Hankel model in Fig.~\ref{fig1}}
\begin{tabular}{|c|c|c|}
\hline
\textbf{Ground truth of poles}& \textbf{Estimated poles with the double Hankel model} & \textbf{Estimated poles with the Hankel model}\\ 
\hline
\rule{0pt}{15pt} $ e^{i2\pi\lbra{0.7568, 0.5952, 0.6019}}$ & $\lbra{1.0000e^{i2\pi0.7583}, \bm{1.0155}e^{i2\pi\bm{0.5970}}, \bm{1.0155}^{-1}e^{i2\pi\bm{0.5970}}}$ &
$\lbra{0.9948e^{i2\pi0.7583}, 1.0466e^{i2\pi0.5951}, 0.9900e^{i2\pi0.3786}}$ \\ \hline
\rule{0pt}{15pt} $ e^{i2\pi\{0.0733, 0.2271, 0.2420\}}$ & $\lbra{1.0000e^{i2\pi0.0730}, \bm{1.0291}e^{i2\pi\bm{0.2331}}, \bm{1.0291}^{-1}e^{i2\pi\bm{0.2331}}}$ &
$\lbra{0.9989e^{i2\pi0.0729}, 0.9739e^{i2\pi0.2241}, 1.0774e^{i2\pi0.2449}}$ \\

\hline
\rule{0pt}{15pt} $ e^{i2\pi\lbra{0.5708, 0.9943, 0.9819}}$&
$\lbra{1.0000e^{i2\pi0.2233}, \bm{1.0331}e^{i2\pi\bm{0.9862}}, \bm{1.0331}^{-1}e^{i2\pi\bm{0.9862}}}$&
$\lbra{0.9501e^{i2\pi0.7540}, 0.9604e^{i2\pi0.9883}, 1.0352e^{i2\pi0.9831}}$ \\
\hline
\end{tabular}
\label{tab1}
\end{table*}

\begin{figure}[htbp]
\centerline{\includegraphics[height=7cm,width=9cm]{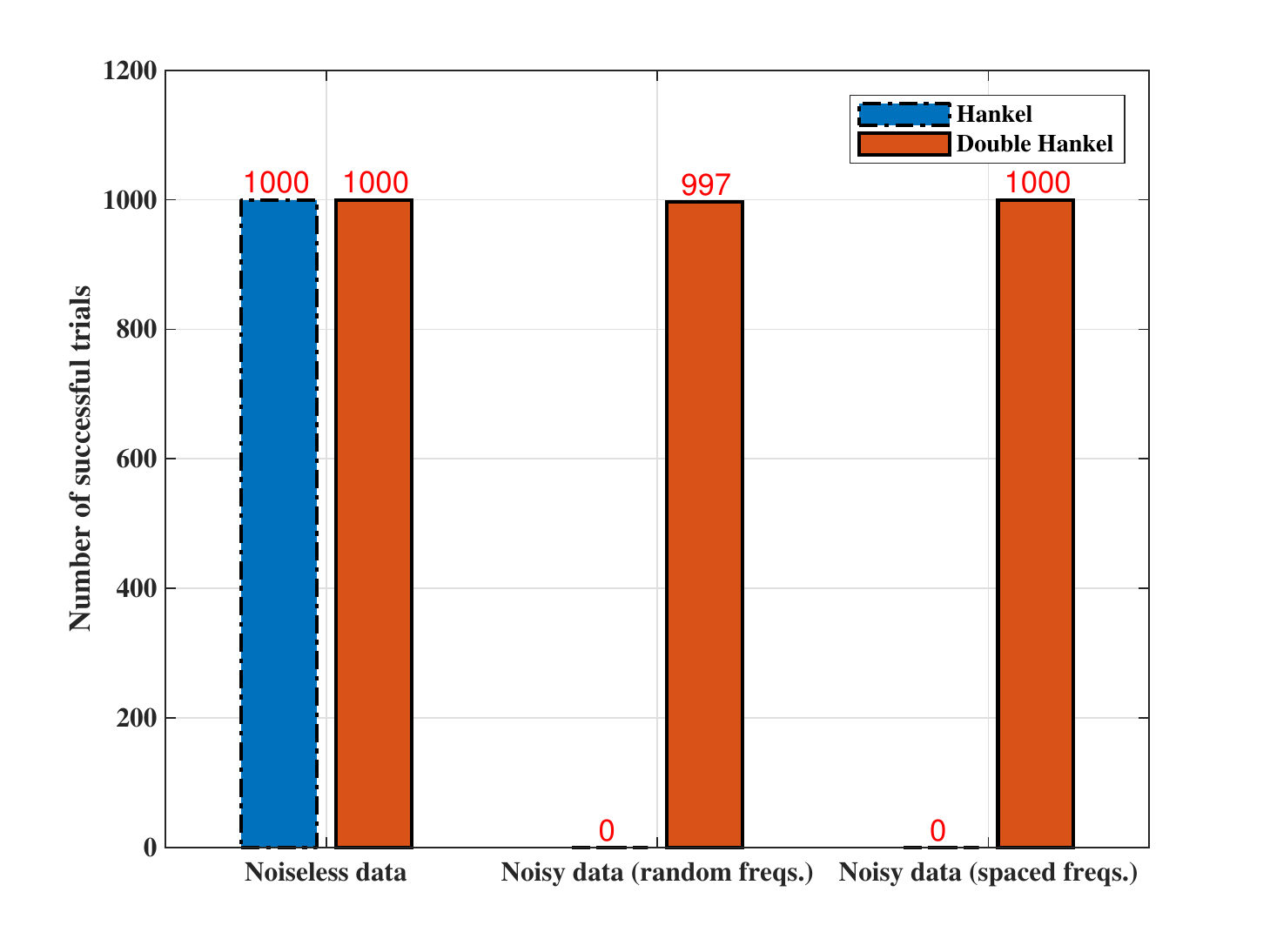}}
\caption{Histogram of the number of successful trials of IHT with the Hankel and double Hankel models with $\text{SNR}=\infty$ and $0$dB and a total number of 1000 trials.}
\label{fig1}
\end{figure}

\begin{figure}[htb]
	\centering
	\subfigure[] {\includegraphics[width=4.2cm]{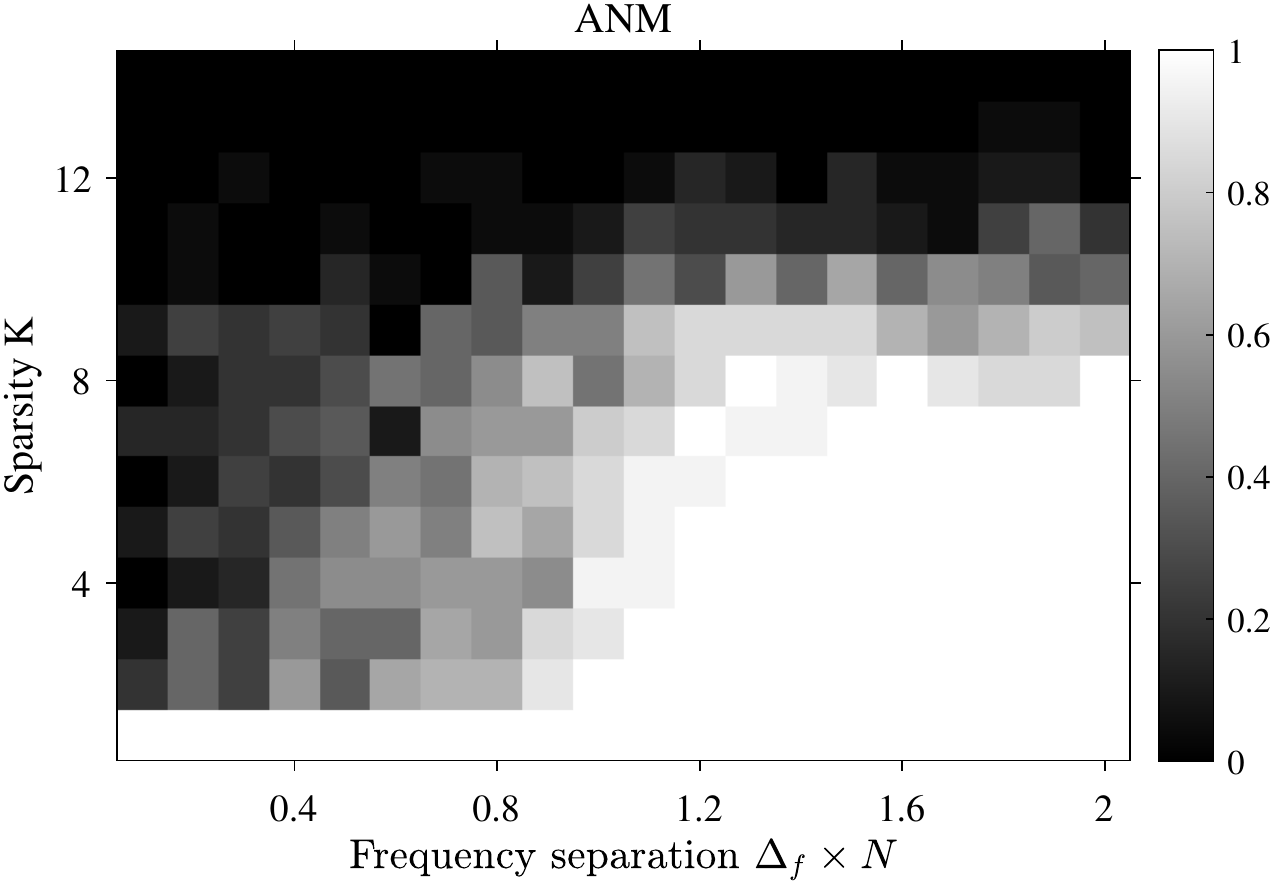}} \\
	\subfigure[] {\includegraphics[width=4.2cm]{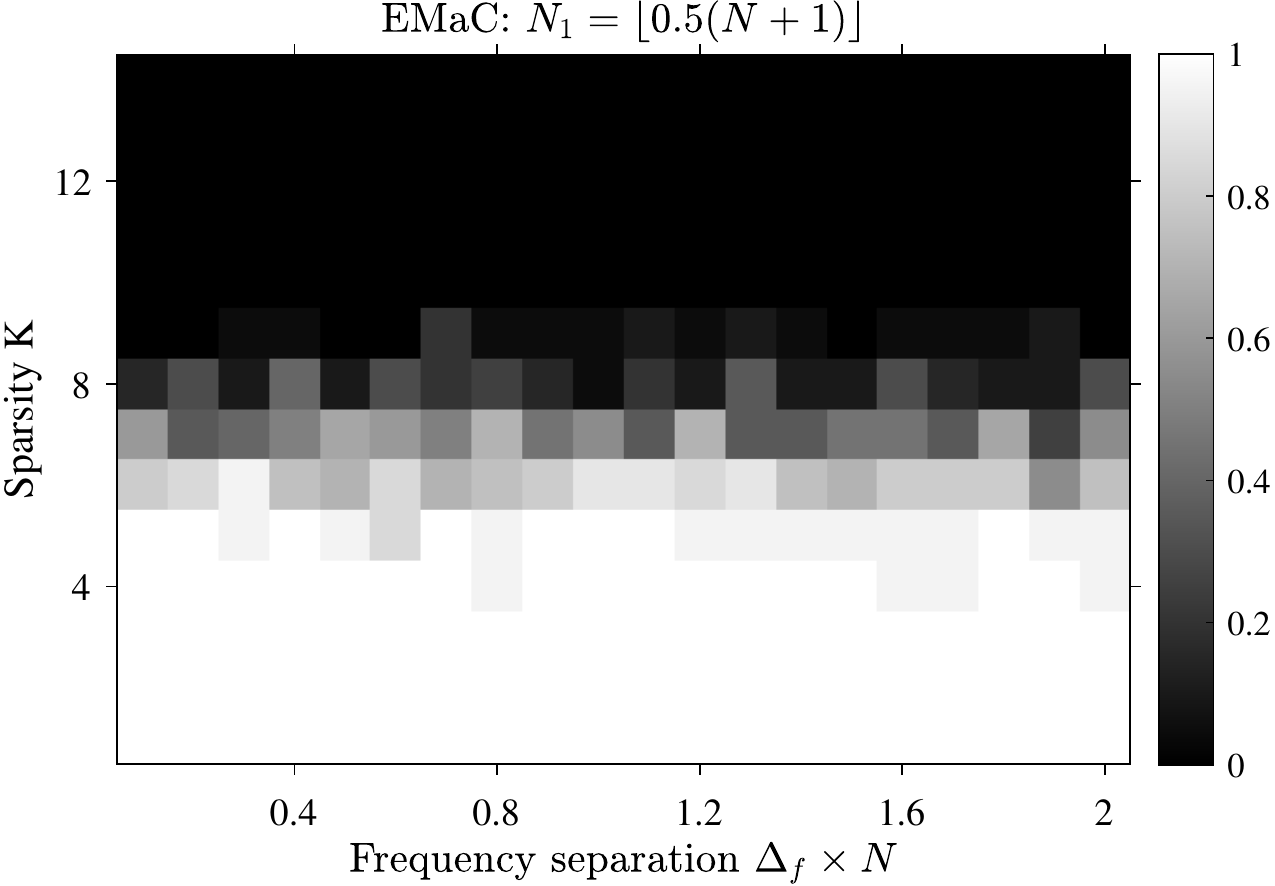}}
	\subfigure[] {\includegraphics[width=4.2cm]{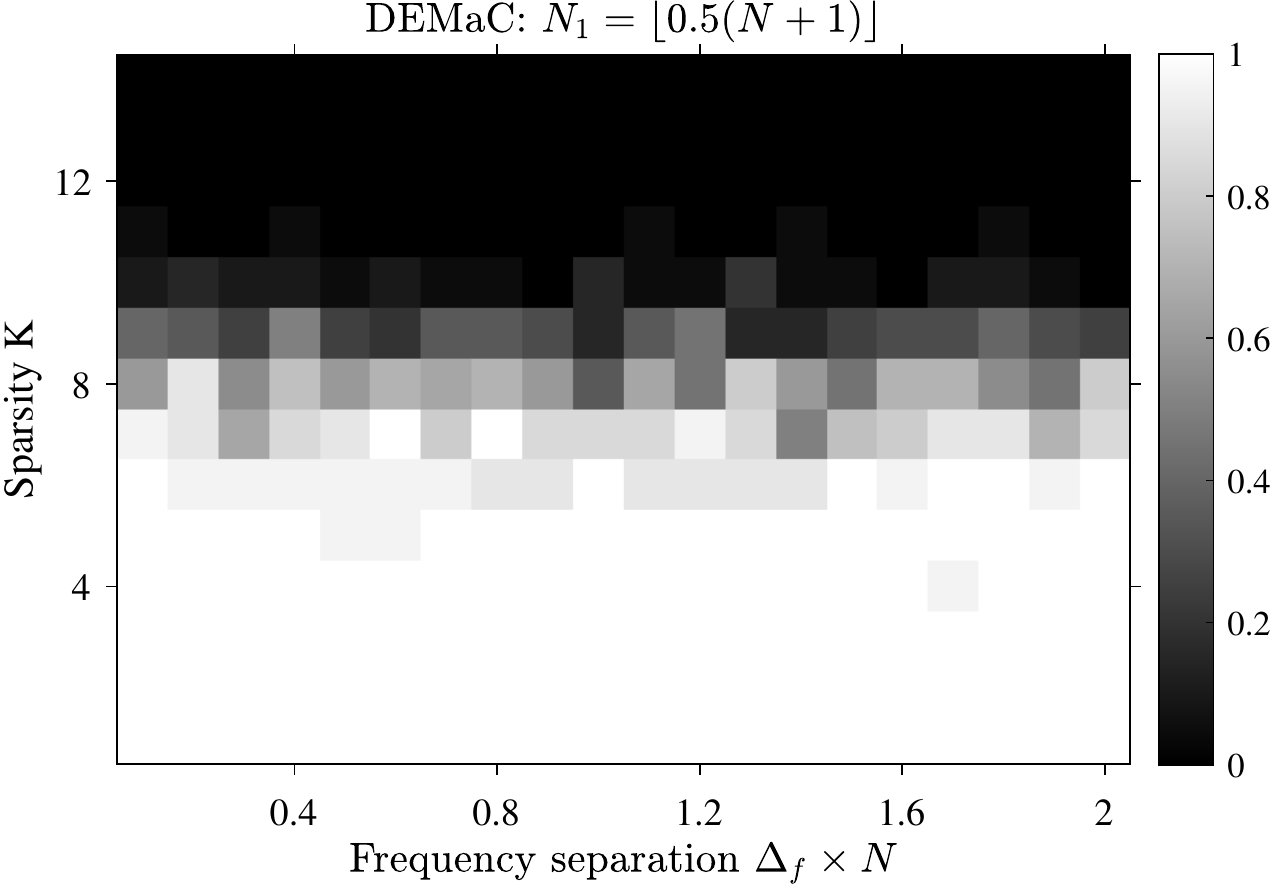}} \\
	\subfigure[] {\includegraphics[width=4.2cm]{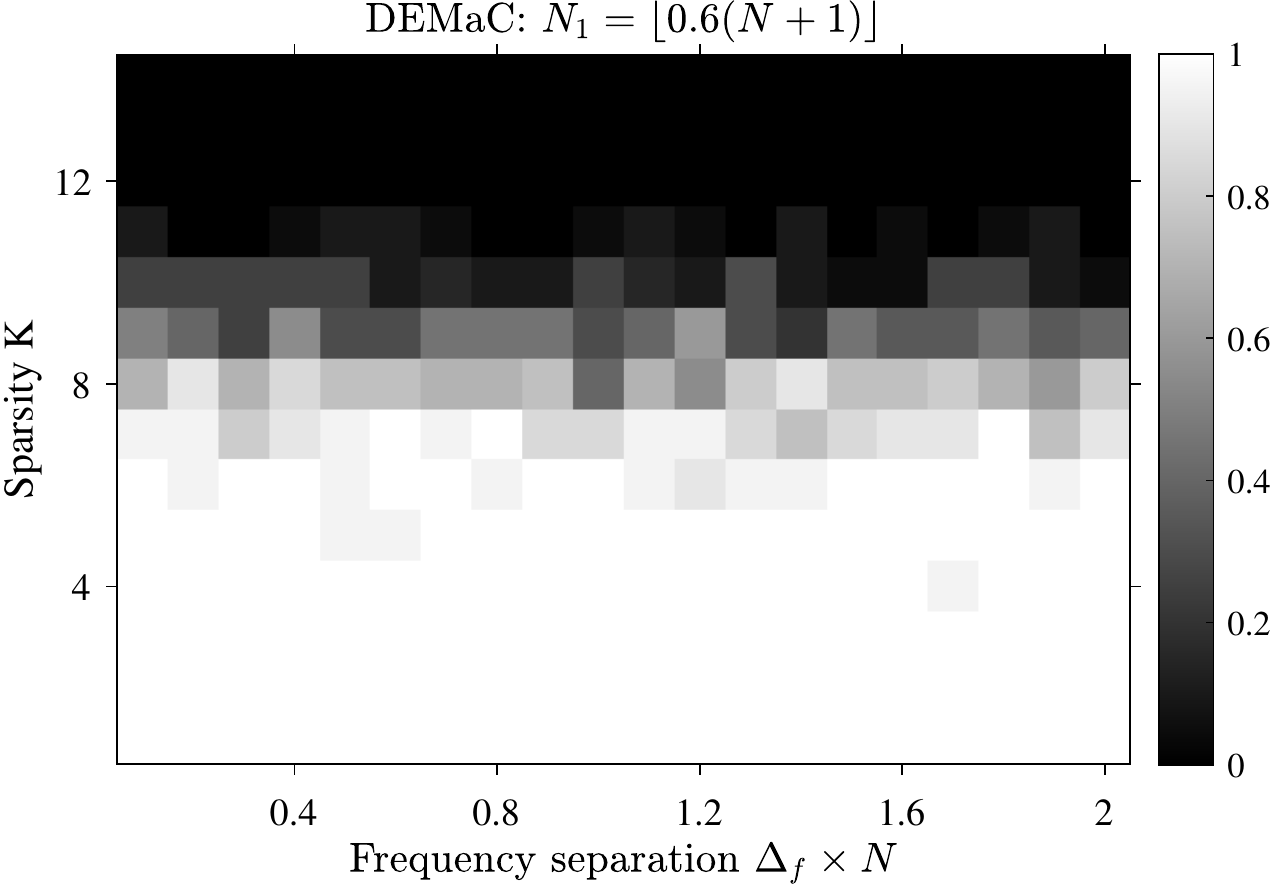}}
	\subfigure[] {\includegraphics[width=4.2cm]{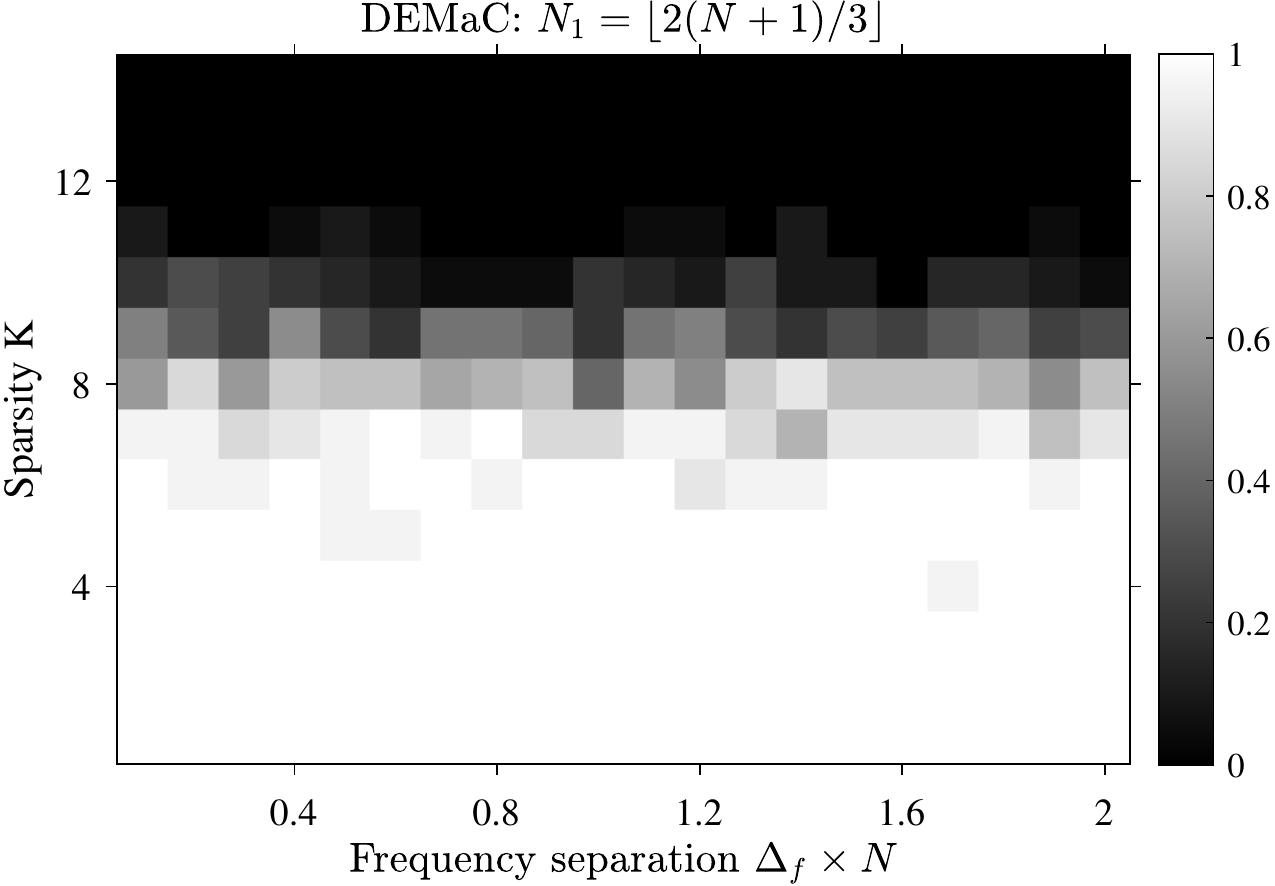}}
	\caption{Sparsity-separation phase transition results of (a) ANM, (b) EMaC with $N_{1}=\left \lfloor 0.5(N+1) \right \rfloor$, and (c)-(e) DEMaC with $ N_{1}=\left \lfloor 0.5(N+1) \right \rfloor, N_{1}=\left \lfloor 0.6(N+1) \right \rfloor$ and $N_{1}=\left \lfloor 2(N+1)/3 \right \rfloor$ with $N=65$ and $M=30$. White means complete success and black means complete failure.}
	\label{Fig-Sparsity-Separation} 
\end{figure}

In our simulation, we consider $K=3$ spectral frequencies that are randomly generated without or with a minimum separation $\frac{4}{N}$ (referred to as \emph{random} and \emph{spaced} frequencies respectively), with amplitudes $0.5+\abs{w}$ and random phases, where $w$ follows the standard normal distribution. A number of $N=65$ uniform samples are acquired with random noise added that follows a complex normal distribution. We set $N_1=N_2=33$ in the Hankel and double Hankel models. The IHT algorithm is terminated if $\norm{\m{y}_{t+1}-\m{y}_{t}}_{2} / \norm{\m{y}_{t}}_{2} < 10^{-5}$ or maximally 3000 iterations are reached. We say that an algorithm successfully produces poles $\lbra{\hat z_k}$ on the unit circle in a trial if their average distance to the unit circle $\frac{1}{K} \sum_{k=1}^{K}{\abs{\abs{\hat z_{k}}-1}}<10^{-4}$.

In our simulation, we consider the noiseless case and the noisy case with the signal to noise ratio $\text{SNR}=0$dB and run 1000 trials for each. The number of successful trials in which the estimated poles lie on the unit circle is shown in Fig.~\ref{fig1}. In the absence of noise, the true spectral poles can be obtained with either Hankel or double Hankel model, as predicted in Theorem \ref{Thm1}. In the presence of noise, the estimated spectral poles with the Hankel model lie off the unit circle, which is consistent with our analysis. In contrast to this, the proposed double Hankel approach rarely fails. It is seen that when the frequencies are properly separated, the IHT with double Hankel model succeeds in all trials to restrict the estimated poles on the unit circle.

To study what happens when the double Hankel model fails to produce poles on the unit circle, we present in Table \ref{tab1} the estimated poles of the three failures shown in Fig.~\ref{fig1}. It is seen that all the failures occur when two frequencies are close to each other with distance less than $\frac{1}{N}$. Their estimated poles are located symmetrically on two sides of the unit circle, which verifies our analysis carried out in Section \ref{sec:FB-Hankel}.

%

\subsection{Accuracy and Robustness of DEMaC}
In this subsection, we study numerical performance of the proposed DEMaC as compared to EMaC \cite{chen2014robust} and atomic norm minimization (ANM) \cite{tang2013compressed}. All three approaches are implemented in Matlab with the CVX toolbox and the SDPT3 solver \cite{grant2008cvx,toh1999sdpt3}. Note that ANM always produces spectral poles on the unit circle and EMaC does not. As in the previous simulation, it is observed that DEMaC produces spectral poles on the unit circle in almost all trials due to the double Hankel model adopted. We next focus on the resolution and accuracy of the three approaches.

We first consider the noiseless case and study the success rate of DEMaC as compared to ANM and EMaC. In particular, a number of $M = 30$ samples are randomly selected from $N = 65$ regularly spaced samples. The coefficients $\{s_{k}\}$ and noise are generated as in the previous simulation. A number of $K$ frequencies are randomly generated such that they are mutually separated by at least $\Delta_{f}$ and two of them are separated exactly by $\Delta_{f}$, where we consider $K\in\lbra{1,2\dots,20}$ and $\Delta_{f}\in\lbra{0,\frac{0.1}{N},...,\frac{2}{N}}$. We say that the spectrally sparse signal $\m{y}^o$ (and its frequencies) is successfully recovered by an algorithm if the normalized mean squared error (NMSE) $\frac{\norm{\hat{\m{y}}-\m{y}^o}_2^2}{\norm{\m{y}^o}_{2}^2}\leq 10^{-10}$. The success rate is calculated by averaging over 20 Monte Carlo trials for each combination $(K,\Delta_{f})$.


Our results are presented in Fig.~\ref{Fig-Sparsity-Separation}, where white means complete success and black means complete failure. Phase transition behaviors are observed in the sparsity-separation plane. It is seen that ANM has a resolution of about $1/N$, as reported in \cite{tang2013compressed}. EMaC is insensitive to the frequency separation, as observed in \cite{chen2014robust}. Like EMaC, DEMaC is insensitive to the separation and has an enlarged success phase thanks to the double Hankel model adopted.

We also present in Fig.~\ref{Fig-Sparsity-Separation} the results of DEMaC with two other choices of $N_1$. It is seen that slightly better performances can be obtained in this case with $N_1\approx 0.6N$ and $N_1\approx 2/3N$. We recommend to use $N_1\approx 0.6N$ in practice.

%
%

\begin{figure}[htb]
	\centering
	\subfigure[Signal recovery error] {\label{Fig-boundednoise-signal} \includegraphics[height=7cm,width=9cm]{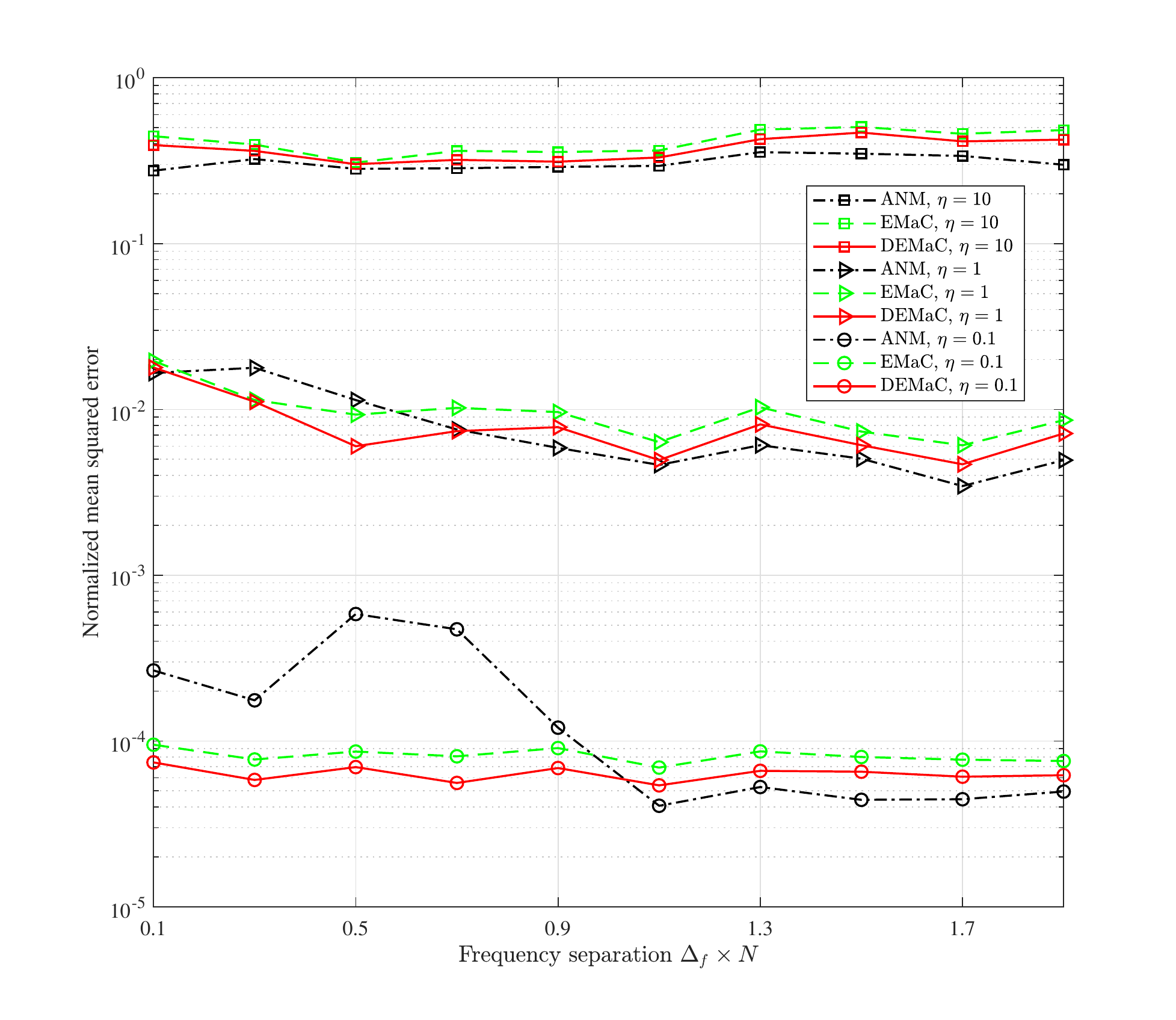}}
\subfigure[Frequency recovery error] {\label{Fig-boundednoise-freq} \includegraphics[height=7cm,width=9cm]{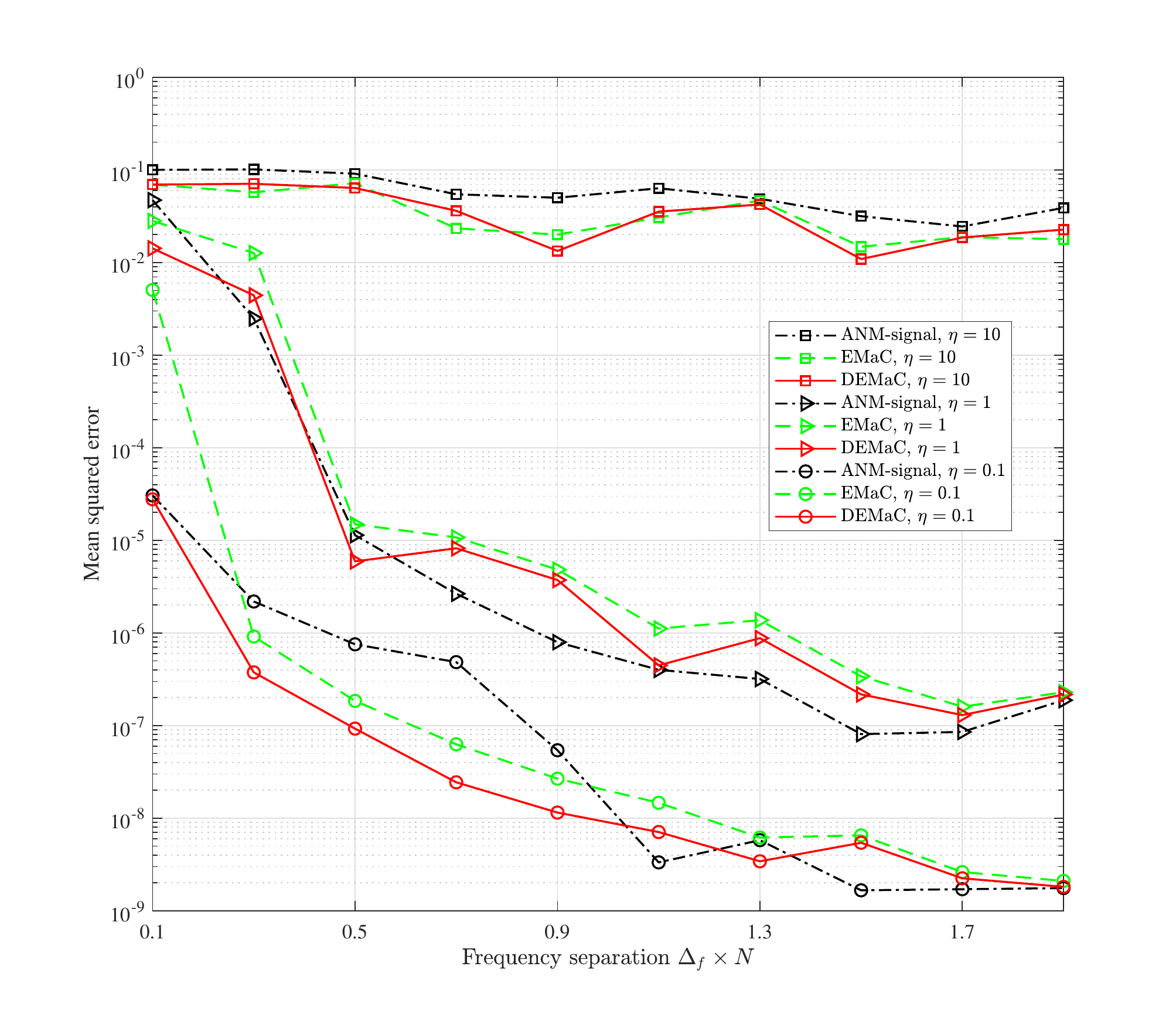}}
\caption{Results of signal and frequency recovery errors using DEMaC, EMaC and ANM with varying frequency separation and noise levels.}
\end{figure}

We test the performance of DEMaC with bounded noise. We consider $K=2$ frequencies, vary their distance $\Delta_f\in\lbra{\frac{0.1}{N},\frac{0.3}{N}, \dots,\frac{1.9}{N}}$ and the noise level $\eta\in\lbra{0.1,1,10}$, set $N_{1}=33$ for both Noisy-EMaC and Noisy-DEMaC, and keep other settings. We plot the curves of signal recovery error in Fig.~\ref{Fig-boundednoise-signal}. It is seen that DEMaC and EMaC are insensitive to the distance of frequencies and the former consistently results in smaller errors than the latter. ANM has the largest errors with closely located frequencies and light noise. We also compare the errors of frequency recovery. Since the model order $K$ might not be correctly determined by using either approach, to make a fair comparison, we assume $K$ is given and estimate the frequencies from the signal estimate using ESPRIT assisted with forward-backward processing. We plot the error curves in Fig.~\ref{Fig-boundednoise-freq}. It is seen that DEMaC also outperforms EMaC and has the smallest error in the case of closely located frequencies.

We now evaluate robustness of DEMaC to additive sparse noise. We consider the full sampling case and vary the sparsity level $K$ and the number of corrupted samples. The success rates of Robust-EMaC and Robust-DEMaC are presented in Fig.\ref{Fig-SparseNoise}. It is seen that DEMaC has an enlarged success phase than EMaC and thus a better performance.

\begin{figure}[htb]
	\centering
	\subfigure[EMaC] {\includegraphics[width=4.2cm]{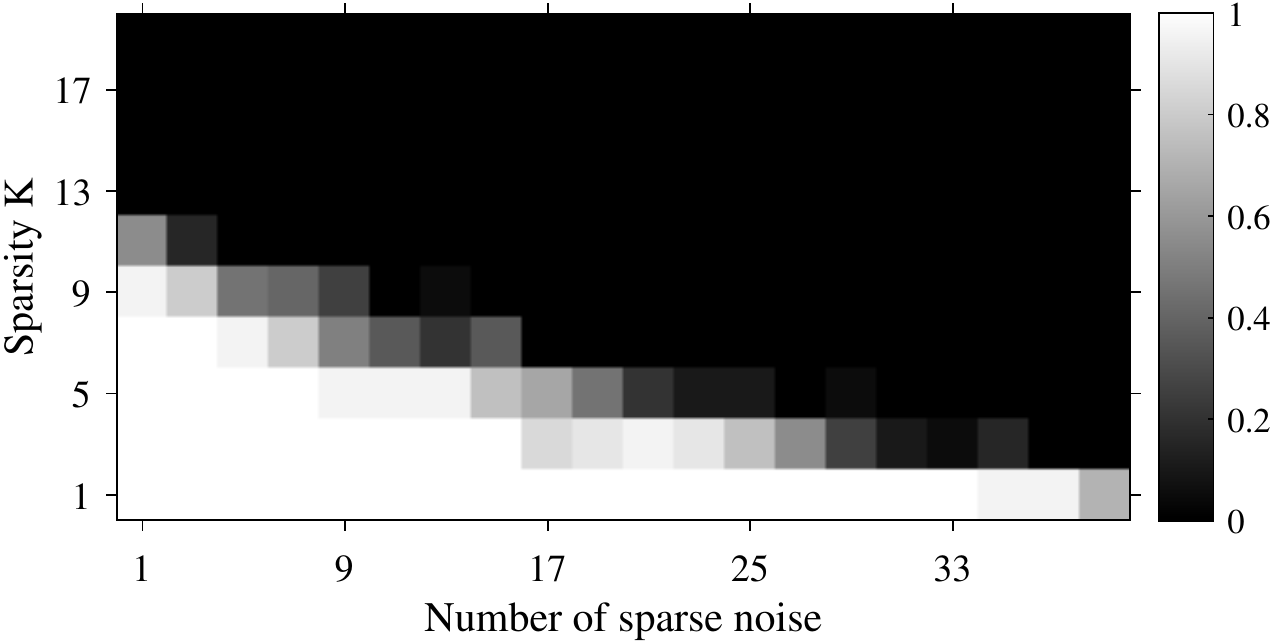}}
	\subfigure[DEMaC] {\includegraphics[width=4.2cm]{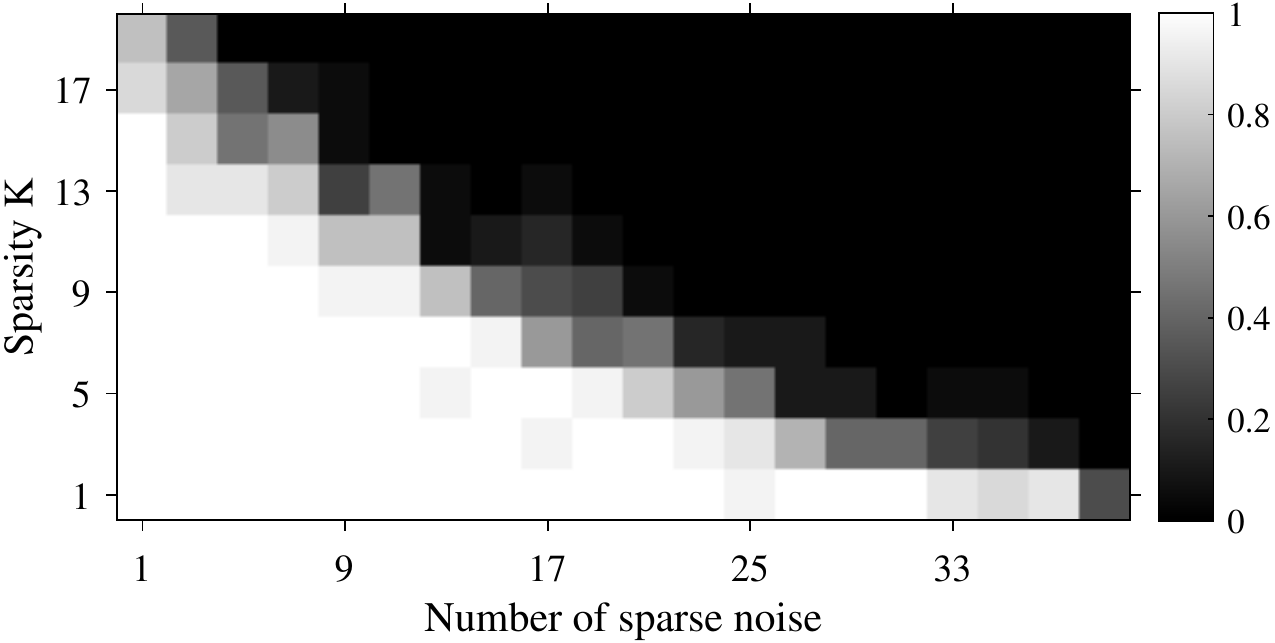}}
	\caption{Results of success rates of (a) EMaC and (b) DEMaC with varying sparsity $K$ and varying number of sparse noise when $N=65$ regularly spaced samples are acquired.}
	\label{Fig-SparseNoise}
\end{figure}

%

\subsection{The 2-D Case}
We present numerical results in the 2-D case in this subsection. We consider recovery of 2-D spectrally sparse signals consisting of $K=3$ spectral components, whose frequencies are randomly generated, from a subset of $N_1\times N_2=11\times 11$ regularly spaced noisy samples. We fix the noise level $\eta=1$, vary the sample size $M\in\lbra{20,24,\dots,120}$ and let $N_{1,1}=N_{1,2}=N_{2,1}=N_{2,2}=6$ for both EMaC and DEMaC. We plot in Fig.~\ref{Fig-2DSignalNMSE_randomseparation} the curves of signal recovery error averaged over 20 Monte Carlo trials. Again, it is seen that DEMaC consistently outperforms EMaC. We also compute the average distance from the estimated signal poles $\lbra{\sbra{\hat{z}_{k,1},\hat z_{k,2}}}$ to the 2-D torus as $\sqrt{\frac{1}{K}\sum_{k=1}^K\sum_{l=1}^2\sbra{\abs{\hat{z}_{k,l}}-1}^2}$ for each trial. The histogram plots for EMaC and DEMaC are presented in Fig.~\ref{Fig-2DDistanceHistogram}. It is seen that the estimated spectral poles of DEMaC are pushed to the 2-D torus (within numerical precision) in most trials due to the double Hankel model adopted.

\begin{figure}[htbp]	\centerline{\includegraphics[height=7cm,width=9cm]{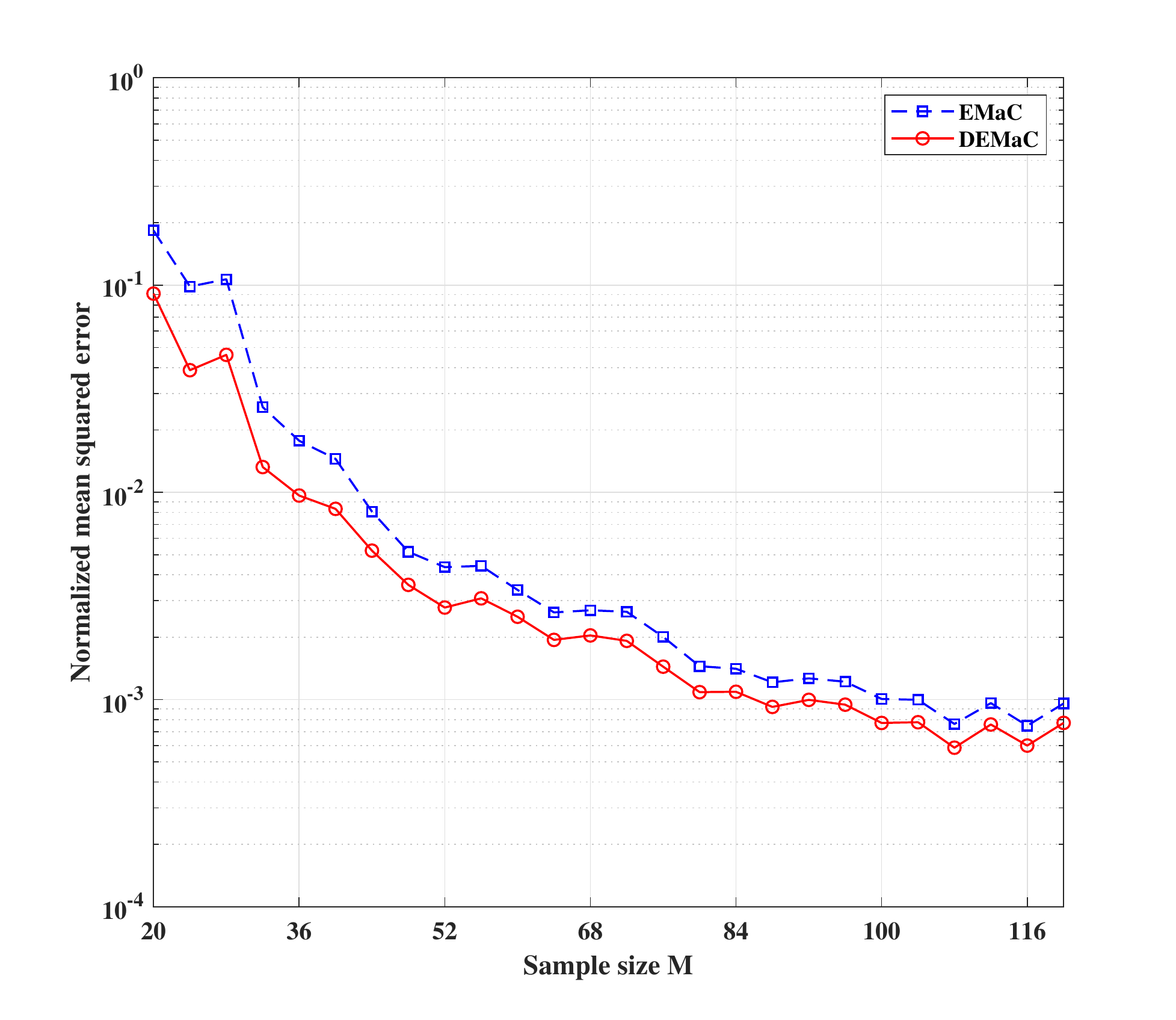}}
	\caption{Results of signal recovery errors of EMaC and DEMaC in the 2-D case with compressive samples and random frequencies.}
	\label{Fig-2DSignalNMSE_randomseparation}
\end{figure}


\begin{figure}[htbp] \centerline{\includegraphics[height=7cm,width=9cm]{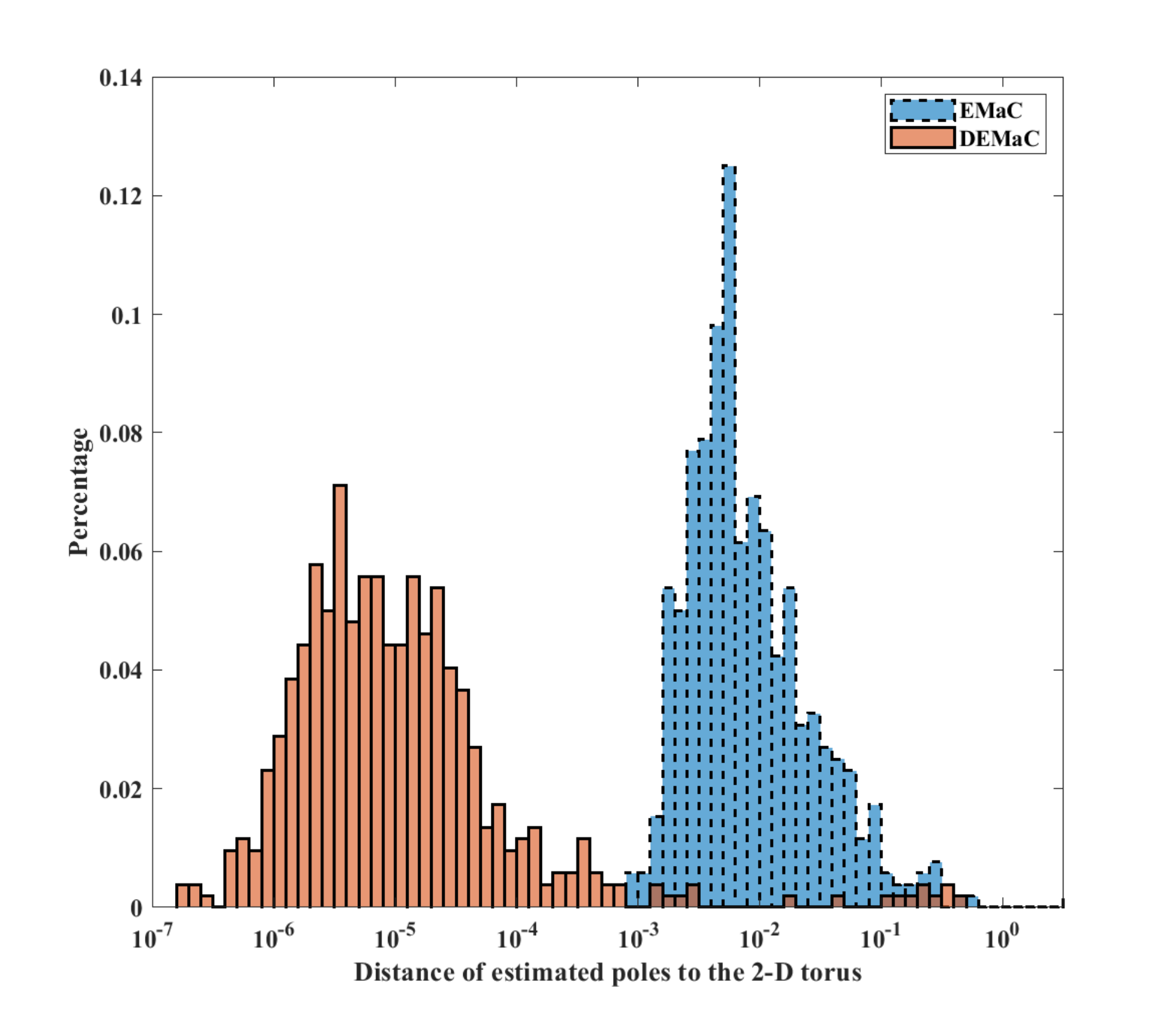}}
\caption{Histogram of distance of the estimated poles to the 2-D torus using EMaC and DEMaC in the 2-D case with compressive samples and random frequencies.}
	\label{Fig-2DDistanceHistogram}
\end{figure}
%

\section{Conclusion} \label{sec:conclusion}
In this paper, a low rank double Hankel model was proposed for arbitrary-dimensional spectral compressed sensing that is capable of pushing the spectral poles to the unit circle and resolves a fundamental limitation of the previous low rank Hankel model. By applying the new model, the convex relaxation DEMaC approaches were proposed that have provable accuracy and are shown theoretically and numerically to outperform EMaC rooted on the Hankel model.

By modifying the IHT algorithm in \cite{cai2019fast}, we have presented in this paper a nonconvex algorithm for spectral compressed sensing with the double Hankel model. A future research direction is to analyze its theoretical performance such as global convergence and accuracy. In fact, it is expected that the double Hankel model can be incorporated into any approach rooted on the Hankel model and brings improvement for recovery of undamped spectrally sparse signals and their spectral poles.

Spectral poles that are located symmetrically on two sides of the unit circle may still appear by using the double Hankel model. Another future research direction is to study new models that exclude such exceptions and meanwhile preserve low model complexity.

%

\appendix

We prove Theorem \ref{thm:noiseless} in this appendix. Our proof is similar to that of \cite[Theorem 1]{chen2014robust} and both are inspired by \cite{gross2011recovering}. We will follow the same steps as in \cite{chen2014robust,gross2011recovering}: 1) identify the sampling basis with respect to the low rank matrix that needs to recover and reformulate DEMaC as affine nuclear norm minimization (note that not all constraints are affine in our case), 2) identify a dual certificate that validates optimality of a solution, 3) show that the sampling basis is incoherent with the tangent space of the ground truth low rank matrix, and 4) construct a dual certificate that validates optimality of the ground truth as a solution to the affine nuclear norm minimization.

\subsection{Problem Reformulation}
As in \cite{chen2014robust}, we try to find the sampling basis and rewrite DEMaC as an affine constrained nuclear norm minimization problem with respect to the low rank matrix $\m{Y} = \mbra{\cH\m{y}\,|\, \m{J}_1\overline{\cH\m{y}}\m{J}_2}$. Note that each sample $y_j^o$, $j\in\Omega$ corresponds to one off-diagonal of each Hankel matrix. Consequently, for every sampling index $n=1,\dots,N$, let $\m{E}_n$ be the (normalized) $n$th $N_1\times N_2$ elementary Hankel matrix with $\m{E}_n\sbra{j,l}=\frac{1}{\sqrt{\omega_n}}$ if $j+l=n+1$ and zeros elsewhere, where $\omega_n$ equals the number of nonzero entries in $\m{E}_n$. Denote by $\cE_n^{(1)}$ (resp.~$\cE_n^{(2)}$) the projection onto the subspace spanned by $\m{E}_n^{(1)}=\mbra{\m{E}_n\,|\,\m{0}}_{N_1\times 2N_2}$ (resp.~$\m{E}_n^{(2)}=\mbra{\m{0}\,|\,\m{E}_{N-n+1}}_{N_1\times 2N_2}$). Then, $\cE_n = \cE_n^{(1)} + \cE_n^{(2)}$ is the $n$th sampling operator with respect to $\m{Y}$. We also define the projection $\cE = \sum_{n=1}^N \cE_n$ and its orthogonal complement $\cE^{\perp} = \cI - \cE$.

For ease of analysis, as in \cite{chen2014robust,gross2011recovering}, we assume that the sampling indices in $\Omega$ are i.i.d.~generated so that $\Omega$ is a multi-set that possibly contains repeated elements. Note that the derived low bounds on the sample size also apply in the case when $\Omega$ is selected uniformly at random from those with distinct elements according to discussions in \cite{gross2011recovering}. Consequently, the sampling operator is given by $\cE_{\Omega}=\sum_{n\in\Omega}\cE_n$. We also define the projection $\cE'_{\Omega}$ that takes sum only over non-repetitive elements in $\Omega$, and its complement operator $\cE_{{\Omega}^{\perp}}' = \cE - \cE'_{\Omega}$.

Using these notations, the Hankel structures in $\m{Y}$ are captured by $\cE^{\perp}\sbra{\m{Y}} = \m{0}$ and the samples of $\m{Y}$ are given by $\cE'_{\Omega}\sbra{\m{Y}} = \cE'_{\Omega}\sbra{\m{Y}^o}$, where $\m{Y}^o=\mbra{\cH\m{y}^o\,|\,\m{J}_1\overline{\cH \m{y}^o}\m{J}_2}$. Note however that the complex conjugate relationship between the two Hankel matrices in $\m{Y}$ is not algebraic and cannot be represented by an affine constraint. To include this constraint, we further define the set
\equ{\cC = \lbra{\mbra{\m{X} \,|\, \m{J}_1\overline{\m{X}}\m{J}_2}:\; \m{X} \in\bC^{N_1\times N_2}}}
and impose $\m{Y}\in \cC$. Now DEMaC is readily rewritten as
\equ{\begin{split}
&\min_{\m{Y}\in\cC} \norm{\m{Y}}_{\star} \\
&\st \cE'_{\Omega}\sbra{\m{Y}} = \cE'_{\Omega}\sbra{\m{Y}^o},\\
&\phantom{\st}  \cE^{\perp}\sbra{\m{Y}} = \cE^{\perp}\sbra{\m{Y}^o} = 0. \end{split} \label{eq:demac1}}
Note that \eqref{eq:demac1} is not a common affine constrained nuclear norm minimization problem due to the restriction onto the feasible set $\cC$.

\subsection{Dual Certificate}
Recall the SVD of $\m{Y}^o$ in \eqref{eq:SVD} and denote by $T$ the tangent space with respect to $\m{Y}^o$. Let $\cP_{\m{U}}$ (resp.~$\cP_{\m{V}}$, $\cP_{T}$) be the orthogonal projection onto the row (resp.~column, tangent) space of $\m{Y}^o$, yielding for any matrix $\m{Y}$ that
\lentwo{\equa{\cP_{\m{U}}\sbra{\m{Y}}
&=& \m{U}\m{U}^H \m{Y}, \\ \cP_{\m{V}}\sbra{\m{Y}}
&=& \m{Y}\m{V}\m{V}^H, \\ \text{and }\cP_{T}
&=& \cP_{\m{U}} + \cP_{\m{V}} - \cP_{\m{U}}\cP_{\m{V}}. }
We denote by $\cP_T^{\perp} = \cI - \cP_T$ the orthogonal complement of $\cP_T$, where $\cI$ is the identity operator.

We provide the dual certificate for DEMaC in the following lemma that is in parallel with \cite[Lemma 1]{chen2014robust}. The proof is almost unaltered and thus is omitted.
\begin{lem}  Consider a multi-set $\Omega$ that contains $M$ random indices. Suppose that the sampling operator $\cE_{\Omega}$ obeys
\equ{\left\|\mathcal{P}_{T} \cE \mathcal{P}_{T}-\frac{N}{M} \mathcal{P}_{T} \cE_{\Omega} \mathcal{P}_{T}\right\| \leq \frac{1}{2}. \label{eq:incoherencecond}}
If there exists matrix $\m{W}$ satisfying
\lentwo{\equa{\cE'_{{\Omega}^{\perp}}(\boldsymbol{W})
&=& \m{0}, \label{eq:Wcons1} \\ \left\|\mathcal{P}_{T}\left(\boldsymbol{W}-\boldsymbol{U} \boldsymbol{V}^{H}\right)\right\|_{\mathrm{F}}
&\leq& \frac{1}{2 N^2}, \label{eq:Wcons2} \\ \left\|\mathcal{P}_{T}^{\perp}(\boldsymbol{W})\right\| &\leq& \frac{1}{2}, \label{eq:Wcons3}
}}then $\m{Y}^o$ is the unique solution to \eqref{eq:demac} or, equivalently, $\m{y}^o$ is the unique minimizer of DEMaC. \label{lem:dualcert}
\end{lem}

\subsection{Proof of Incoherence Condition \eqref{eq:incoherencecond}}
To show the incoherence condition \eqref{eq:incoherencecond}, we first bound the projection of each $\m{E}_n^{\sbra{j}}$ onto the tangent space $T$ in the following lemma that is in parallel with \cite[Lemma 2]{chen2014robust}. The proof is similar and will be omitted.
\begin{lem} Given \eqref{eq:incoh}, we have
\equ{\cP_{\m{U}}\sbra{\m{E}_n^{\sbra{j}}}\leq \frac{\mu_1c_sK}{N},\quad \cP_{\m{V}}\sbra{\m{E}_n^{\sbra{j}}}\leq \frac{\mu_1c_sK}{N}}
for all $n=1,\dots,N$ and $j=1,2$. For any $n_1,n_2=1,\dots,N$ and any $j_1,j_2=1,2$, we have
\equ{\abs{\inp{\m{E}_{n_1}^{\sbra{j_1}},\; \cP_T\sbra{\m{E}_{n_2}^{\sbra{j_2}}}}} \leq \sqrt{\frac{\omega_{n_2}}{\omega_{n_1}} } \frac{2\mu_1c_sK}{N}.} \label{lem:PTE}
\end{lem}

The incoherence condition \eqref{eq:incoherencecond} is established in the following lemma that is in parallel with \cite[Lemma 3]{chen2014robust}.

\begin{lem} For any constant $0<\epsilon\leq\frac{1}{2}$, we have
\equ{\left\|\mathcal{P}_{T} \cE \mathcal{P}_{T}-\frac{N}{M} \mathcal{P}_{T} \cE_{\Omega} \mathcal{P}_{T}\right\| \leq \epsilon \label{eq:incoherencecond2}}
with probablity exceeding $1-N^{-4}$, provided that $M> c_1\mu_1c_sK\log N$ for some universal constant $c_1>0$.
\end{lem}

\begin{proof} The proof is similar to that of \cite[Lemma 3]{chen2014robust} and we mainly highlight a few differences. We define a family of operators
\equ{\cZ_n = \frac{N}{M}\cP_T\cE_n\cP_T - \frac{1}{M}\cP_T\cE\cP_T}
for $n=1,\dots,N$. It is seen that
\equ{\mathcal{P}_{T} \cE \mathcal{P}_{T}-\frac{N}{M} \mathcal{P}_{T} \cE_{\Omega} \mathcal{P}_{T} = -\sum_{n\in\Omega} \cZ_n.}
As in \cite{chen2014robust}, we can compute
\equ{\norm{\cP_T\cE_n^{(j)} \cP_T} \leq \frobn{\cP_T\sbra{\m{E}_n^{(j)} }}^2 \leq \frac{2\mu_1c_sK}{N},}
where the second inequality follows from Lemma \ref{lem:PTE}. Thus,
\equ{\norm{\cP_T\cE_n \cP_T} \leq \sum_{j=1}^2\norm{\cP_T\cE_n^{(j)} \cP_T} \leq \frac{4\mu_1c_sK}{N}.}

For any $n\in\Omega$ that is uniformly drawn from $\lbra{1,\dots,N}$, by derivations similar to those in \cite{chen2014robust}, we have
\lentwo{\equa{ \bE\mbra{\cZ_n}
&=& 0, \\ \norm{\cZ_n}
&\leq& 2\max_{n} \frac{N}{M} \norm{\cP_T\cE_n \cP_T}\leq \frac{8\mu_1c_sK}{M}, \label{eq:znnorm}\\ \sum_{n\in\Omega}\norm{\bE\mbra{\cZ_n^2}}
&\leq& \frac{N}{M}\norm{\cP_T\cE_n \cP_T} +\frac{1}{M} \leq \frac{8\mu_1c_sK}{M}. \label{eq:znsnorm}
}}The conclusion is finally drawn by also applying the Bernstein inequality \cite[Theorem 1.6]{tropp2012user} (see also \cite[Lemma 11]{chen2014robust}).

It is worth noting that the upper bounds in \eqref{eq:znnorm} and \eqref{eq:znsnorm} are amplified by a scaling factor 2 as compared to those in \cite{chen2014robust}. Consequently, the universal constant $c_1$ is doubled.
\end{proof}

\subsection{Construction of Dual Certificate $\m{W}$}
We construct a dual certificate $\m{W}$ using the golfing scheme introduced in \cite{gross2011recovering} as in \cite{chen2014robust}. In particular, suppose that $\Omega = \cup_{i=1}^{j_0} \Omega_{i}$ where $\Omega_i$'s are independently generated multi-sets, each containing $\frac{M}{j_0}$ i.i.d.~samples. Let $\epsilon<\frac{1}{e}$, $j_0=5\log_{\frac{1}{\epsilon}}N$ and $q = \frac{M}{Nj_0}$. The dual certificate $\m{W}$ is constructed in the following three steps:
\begin{enumerate}
 \item Set $\m{F}_0=\m{U}\m{V}^H$;
 \item For all $i=1,\dots,j_0$, let $\m{F}=\cP_T\sbra{\cE-\frac{1}{q}\cE_{\Omega_i}}\cP_T\sbra{\m{F}_{i-1}}$;
 \item Set $\m{W} = \sum_{i=1}^{j_0} \sbra{\frac{1}{q}\cE_{\Omega_i}+\cE^{\perp}}\sbra{\m{F}_{i-1}}$.
\end{enumerate}

The remaining task is to show that $\m{W}$ satisfies \eqref{eq:Wcons1}--\eqref{eq:Wcons3} in Lemma \ref{lem:dualcert}. In fact, \eqref{eq:Wcons1} and \eqref{eq:Wcons2} can be shown by the same arguments as in \cite{chen2014robust}. The proof of \eqref{eq:Wcons3} is also similar if we change the definitions of the norms $\norm{\cdot}_{\cE,\infty}$, $\norm{\cdot}_{\cE,2}$ introduced in \cite{chen2014robust} as:
\lentwo{\equa{ \norm{\m{Y}}_{\cE,\infty}
&=& \max_{n}\sum_{j=1}^2\frac{\abs{\inp{\m{E}_n^{(j)},\; \m{Y}}}} {\sqrt{\omega_n}}, \label{eq:Yinfn} \\ \norm{\m{Y}}_{\cE,2}
&=& \sqrt{\sum_{n=1}^N \sum_{j=1}^2 \frac{\abs{\inp{\m{E}_n^{(j)},\; \m{Y}}}^2} {\sqrt{\omega_n}} \label{eq:Y2n}}.
}}Based on \eqref{eq:Yinfn} and \eqref{eq:Y2n}, we can prove analogous versions of Lemmas 4--7 in \cite{chen2014robust} with minor modifications that together result in \eqref{eq:Wcons3}. We will omit the details.

It is worth noting that the constraint $\m{Y}\in\cC$ is not utilized in our proof (so that the number of free variables is doubled) and the resulting universal constant $c_1$ is doubled as compared to that in \cite{chen2014robust}. After all, the derived sample size shares the same order-wise complexity as in \cite{chen2014robust} that is of essential importance in such big-data analysis. It is interesting to investigate in future studies how to use the constraint $\m{Y}\in\cC$ to further reduce the sample size.



\end{document}